\documentclass[aps,prper,showpacs,twocolumn,floats,superscriptaddress,10pt]{revtex4-1}
\usepackage{graphicx}
\usepackage{dcolumn}
\usepackage{bm}
\usepackage{amssymb}
\usepackage{nicefrac}

\usepackage[latin3]{inputenc}
\usepackage[makeroom]{cancel}
\usepackage{amsmath}
\usepackage{amsfonts}
\usepackage{amssymb}
\usepackage{color}
\usepackage[left=2cm,right=2cm,top=2cm,bottom=2cm]{geometry}

\usepackage{graphicx} 
\usepackage{dcolumn}
\usepackage{bm}
\usepackage{simplewick}
\usepackage{array}
\usepackage{appendix}

\RequirePackage[
   hyperindex,colorlinks,bookmarksnumbered,
   plainpages=true,pdfstartview=FitH]{hyperref}
\hypersetup{linkcolor=blue,urlcolor=blue,citecolor=blue}
\usepackage{hyperref}

\definecolor{purple}{rgb}{0.5,0,0.6}

\usepackage[export]{adjustbox}
\usepackage{ulem}

\begin{document}
\title{Thermoelectric transport across a tunnel contact between two charge Kondo circuits{\color{black}: beyond perturbation theory}}
\date{\today}

\author{T. K. T. Nguyen}
\email{nkthanh@iop.vast.vn}
\affiliation{Institute of Physics, Vietnam Academy of Science and Technology, 10 Dao Tan, 118000 Hanoi, Vietnam}
\author{H. Q. Nguyen}
\affiliation{Institute of Physics, Vietnam Academy of Science and Technology, 10 Dao Tan, 118000 Hanoi, Vietnam}
\author{M. N. Kiselev}
\affiliation{The Abdus Salam International Centre for Theoretical Physics, Strada
Costiera 11, I-34151, Trieste, Italy}

\begin{abstract}
Following a theoretical proposal on multi-impurity charge Kondo circuits [T. K. T. Nguyen and M. N. Kiselev, Phys. Rev. B {\bf 97}, 085403 (2018)] and the experimental breakthrough in fabrication of the two-site Kondo simulator [W. Pouse {\it et al}, Nat. Phys. (2023)] we investigate a thermoelectric transport
through a {\color{black} double-dot} charge Kondo quantum nano-device in the strong coupling operational regime.  We focus on the fingerprints of the non-Fermi liquid and its manifestation in the charge and heat quantum transport. We construct a full-fledged quantitative theory describing crossovers between different regimes of the multi-channel charge Kondo quantum circuits and  discuss  possible experimental realizations of the theory.
\end{abstract}

\maketitle

\section{Introduction}

Thermoelectric materials have been investigated in recent years thanks to their ability to generate electricity from waste heat or being used as solid-state Peltier coolers \cite{TE_materials}. The mechanism of converting of heat into voltage known as the Seebeck effect \cite{Seebeck} is associated with the emergence of the electrostatic potential across the {\it hot} and {\it cold} ends of the {\it thermocouple}  \cite{Seebeck1,Seebeck2} while no electric current flows through the system.  The Peltier effect is manifested by  the creation of the temperature difference between the junctions when the electric current flows through the {\it thermocouple}.
 
After theoretical predictions have been suggested that the thermoelectric efficiency could be greatly enhanced through nano-structural engineering in the mid-1990s \cite{Hicks,Mahan}, many complex nano-structured materials were studied in both theory and experiment \cite{lowD_TE_materials,lowD_TE_materials1, lowD_TE_materials2, Blanter,Kisbook}. Nano-electric circuits based on one or a few quantum dots (QDs), which
are highly controllable and fine-tunable, can provide important information about the effects of strong electron-electron interactions, interference effects and resonance scattering on the quantum charge, spin and heat transport.

One of the fundamental motivations of the thermoelectric studies is
to enhance thermoelectric power (absolute value of the Seebeck coefficient, TP). It is a challenge for both experimental fabrication of devices and theoretical suggestions for efficient mechanisms of heat transfer. In fact, {\color{black} many} theoretical investigations showed that the TP of a single electron transistor (SET) was greatly enhanced in comparison with those of bulk materials \cite{staring_93,Turek_Matveev,TE_QD1,TE_QD2,TE_QD3,TE_QD4,TE_QD5} {\color{black} and this has also been realized in experiments \cite{exp_TE_QD1,exp_TE_QD2}}. 
Furthermore, the charge Kondo effect \cite{flensberg,matveev,andreevmatveev,furusakimatveev,LeHur1,LeHur2} dealing with the degeneracy of the charge states of the QD (which is similar to the conventional Kondo effect \cite{Kondo,Hewson,TW1983,AFL1983,Kondo_review} but does not require the system to have magnetic degree of freedom) can be a
tool for intensification of the TP of a SET \cite{andreevmatveev,TP_Kondo_exp}. The building block of a charge Kondo circuit (CKC) is a large metallic QD strongly coupled to one (or several) lead(s) through an (or several) almost transparent single-mode quantum point contact(s) [QPC(s)]. In the {\it orthodoxal} charge Kondo theory \cite{flensberg,matveev,furusakimatveev}, the electron location (namely, in or out of QD) is treated as an iso-spin variable, while two spin projections of electrons
are associated with two (degenerate, in the absence of external magnetic field) conduction channels in the conventional Kondo problem. External magnetic field lifts out the channel degeneracy resulting in a crossover from two channel Kondo (2CK) regime at the vanishing magnetic field to the single channel Kondo (1CK) regime at the strong external field \cite{LeHur2,thanh2010}. As a result, the behavior of the system continuously changes from non-Fermi liquid (NFL) to the Fermi liquid (FL) states respectively. The interplay between NFL-2CK and FL-1CK regimes in thermoelectric transport through the SET has {\color{black} also} been investigated {\color{black} in different situations such as: 
materials with strong spin-orbit interaction \cite{thanh2015}; quantum simulators with strong many-body interactions between mobile carriers \cite{Anton2022,Thanh_VN_3}.  The proposal of the quantum simulators consisting of two weakly coupled SETs  has been suggested in \cite{thanh2018,Thanh_VN_2}. Recently, the challenging problem of a generalized Wiedemann-Franz (WF) law in the quantum simulators and its connection  to the Anderson orthogonality catastrophe has been addressed in  \cite{Kis2023}. Beside, the effects of the electron-electron interactions in the charge Kondo simulators have also been considered \cite{Thanh_VN_1,thanh2023,anton2023}. In those works, the bosonization (and refermionization) method is applied.} The charge transport in 1CK and 2CK regimes were studied extensively numerically in Refs.\cite{Num1,Num2,Num3,Num4}. {\color{black} Developing numerical methods for a quantum impurity problem out of equilibrium is still a big challenge. There is a recent progress on the charge Kondo effects by using quantum Monte Carlo (QMC) technique \cite{QMC1,QMC2,QMC3}. Although the numerical renormalization group \cite{Num1,Num2,Num3,Num4} is more developed and advanced in comparison to the QMC now days, some recent works on QMC on the Keldysh contour \cite{QMC4,QMC5} pave a way for developing new approaches and provide complementary tools for better understanding of both new theoretical models and new experiments with quantum simulators.}

Recently, CKCs operated in the integer quantum Hall (IQH) regime have been implemented in breakthrough experiments \cite{pierre2,pierre3}. With the advantage that the number of Kondo channels is determined by the number of QPCs attached to the metallic QD, these experiments have opened an access to investigation of the multi-channel Kondo (MCK) problem experimentally. The dominant characteristic of a specific MCK setup is a NFL picture \cite{NB1980,Cox1998,AL1993} which
is associated with $Z_{M}$ symmetry. For instance, the NFL-2CK \cite{AD1984,FGN_1,FGN_2}
is explained by Majorana fermions \cite{gogolin,Toulous_limit}, the
NFL-3CK physics is related to $Z_{3}$ parafermions \cite{Z3_1,Z3_2,Z3_3,Z3_4,Z3_5,Z3_6}.
Therefore, switching between $Z_{2k+1}$ and $Z_{2k}$ low temperature
fixed points by controlling the reflection amplitudes of the QPCs,
can provide a route to investigate the crossovers between states with
different parafermion fractionalized zero modes \cite{thanhprl}.

As a CKC is considered as an artificial quantum simulator for the technology of quantum computer, scaling up the CKCs to clusters or lattices is challenging and it is important to understand the nature of the coupling between neighboring QDs. For this motivation, the experiment \cite{Gordon2023} has implemented a two-island charge Kondo device in which two QDs are coupled together and each one is also strongly coupled to an electrode through a QPC. The authors investigated the quantum phase transition at the triple point where the charge configurations are degenerate. Being more than the two-impurity Kondo (2IK) model, the two-site charge Kondo circuit {\color{black}(2SCKC)} is relevant to the Kondo lattice systems. Furthermore, the deeper theoretical investigation of the strong central coupling of this setup \cite{Karki2022,Z3_DCK} in the Toulouse limit showed that a $Z_{3}$ parafermion emerging at the critical point, was already present in the experimental device of Ref. \cite{Gordon2023}.

\begin{figure}
\begin{tabular}{c}
\includegraphics[width=1\columnwidth]{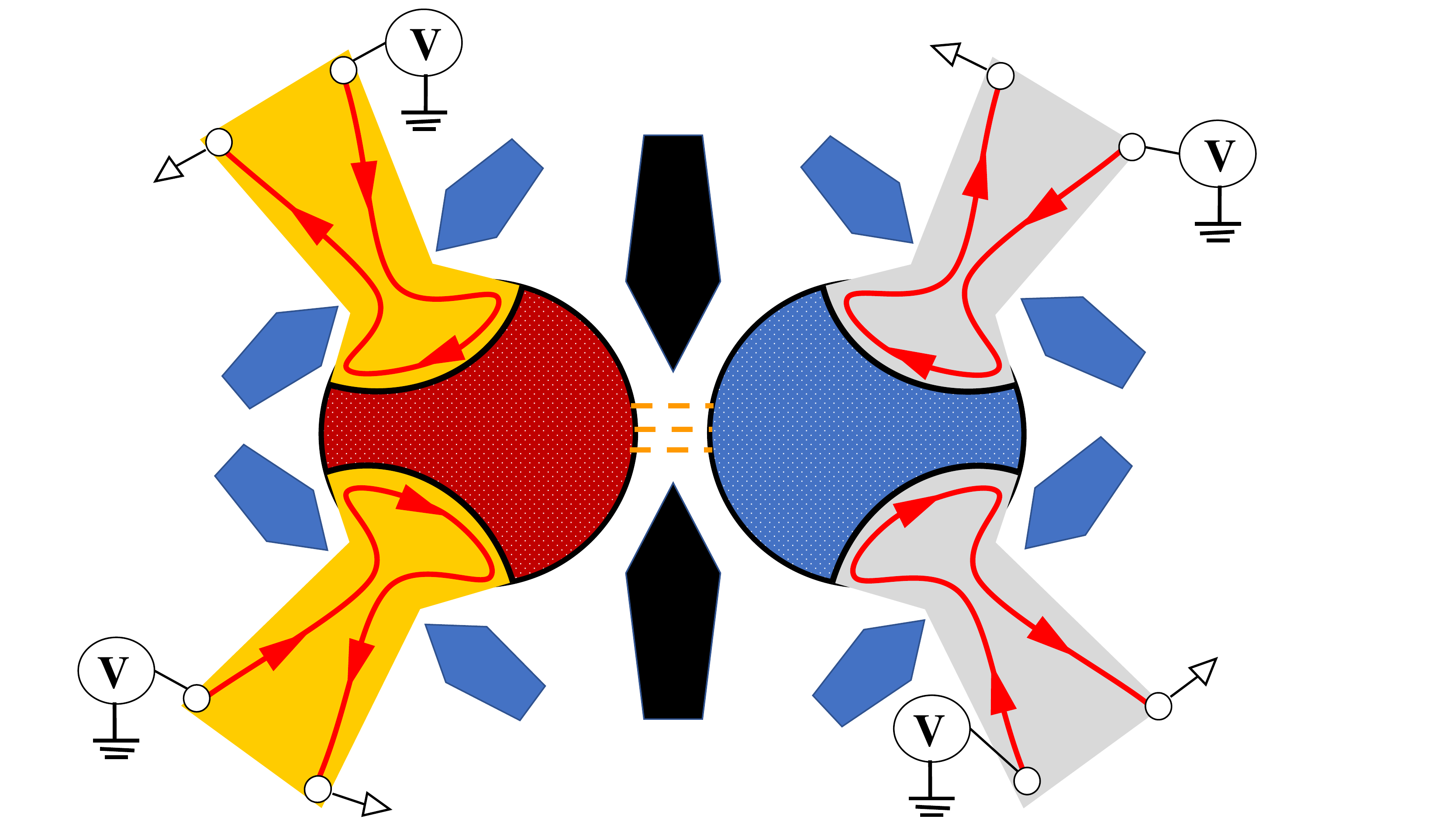}\tabularnewline
\end{tabular}\caption{Schematic of a weak link between two charge Kondo circuits (CKC). Each circuit consists of a large metallic island (QD), which is embedded into two-dimensional electron gas (2DEG) and connects to two large electrodes through the single-mode quantum point contacts (QPCs). The 2DEG (plain area) is in the integer quantum Hall regime $\nu=1$. The red line with arrows denotes the chiral edge mode which backscatters at the center of the narrow constriction. The QPCs are fine tuned by field effects in split gates (blue boxes) to different regimes. We name the QPCs in the left CKC as $\text{QPC}_{11}$, $\text{QPC}_{12}$ and the QPCs in the right CKC as $\text{QPC}_{21}$, $\text{QPC}_{22}$. The right CKC (grey color) is at the reference temperature $T$ while the left circuit (orange color) is at higher temperature $T+\Delta T$.}
\label{Fig1}
\end{figure}

In this work, we revisit the model proposed in Ref. \cite{thanh2018}
which contains a tunnel contact between two CKCs where each one is
set up in either FL or NFL state (see Fig. \ref{Fig1}) with a two-fold goal. First, we examine the behavior of TP in order to find a mechanism to enhance it. Secondly, we show the existence of Majorana fermions in this {\color{black}2SCKC}. 
{\color{black} The approach used in \cite{thanh2018} is based on accounting for the perturbative corrections to the transport off-diagonal coefficients and is limited by the perturbation theory domain of validity (high-temperature regime). These calculations, being very useful for understanding the flow towards the non-Fermi liquid intermediate coupling fixed point, neither become valid at the low-temperature regime, nor shed a light on reduction of the symmetry due to the emergency of the Majorana (parafermionic) states. The main idea of this work is to develop a controllable and reliable approach for the quantitative description of the Fermi-to-non-Fermi liquid crossovers and interplay around the intermediate coupling fixed points. It therefore provides a complementary study of the model  \cite{thanh2018} and completes the theory of thermoelectrics in 2SCKCs.}
The new energy scale associated with the inverse lifetime of the  emergent Majorana fermions controls four different regimes of thermoelectric transport based on the window of parameters. 

The paper is organized as follows. We {\color{black} briefly describe the proposed experimental setup and general equations for the thermoelectric coefficients in Sec. II. The Sec. III represents the correlation function in different cases. The main results are represented in Sec. IV. We discuss the results and conclude our work in Sec. V.}

\section{Proposed experimental setup and general equations for the thermoelectric coefficients}

\subsection{Proposed experimental setup}

We consider a {\color{black}2SCKC} device (see Fig. \ref{Fig1}) formed by two CKCs
describing a very recent experiment \cite{Gordon2023}. The building
block for each CKC is a QD-QPC structure implemented in experiment
\cite{pierre2}. The QD is a large metallic island (the dark-red and
blue cross-hatched areas surrounded by the black lines) electronically connected to a two-dimensional electron gas (2DEG, the orange and
grey continuous areas). The 2DEG is connected to two large electrodes
through two QPCs. Applying a strong magnetic field perpendicular to
the 2DEG plane can control the 2DEG in the IQH regime at the filling
factor $\nu=1$. The QPCs are fine-tuned (by field effects in the split gates illustrated by the blue boxes) to the high transparency regime
corresponding to weak backscattering of the chiral edge mode 
(red solid lines with arrows). We investigate the regime  of equal reflection amplitudes at two QPCs in each CKC:
$|r_{11}|=|r_{12}|=|r_{1}|$ and $|r_{21}|=|r_{22}|=|r_{2}|$.
Therefore, each CKC is a 2CK setup. Indeed, the CKC can be tuned into a 1CK model by simply deactivating one of the two QPCs in it. These two CKCs are connected together by a weak tunneling (barrier, weak link) between two QDs. In order to study the thermoelectric transport through the {\color{black}2SCKC} system, the left CKC is set up at higher temperature $T+\Delta T$ in comparison with the right circuit, which is at temperature $T$. The temperature drops at the central weak link.

{\color{black}
\subsection{General formulas for the thermoelectric coefficients}

In order to study the thermoelectric effects at the weak link between two QDs in the linear response regime $[\Delta T, e\Delta V]\ll T$ (we adopt the units $\hbar$$=$$c$$=$$k_B$$=$$1$), we consider both the charge current $I_e$ and the heat current $I_h$ across the tunnel contact \cite{Onsager}:
\begin{equation}
\left(\begin{array}{c}
I_{e}\\
I_{h}
\end{array}\right)=\left(\begin{array}{cc}
G & G_{T}\\
TG_{T} & G_{H}
\end{array}\right)\left(\begin{array}{c}
\Delta V\\
\Delta T
\end{array}\right),
\end{equation}
where $G, G_T, G_H$ are the electric conductance, thermoelectric coefficient, and thermal coefficient, correspondingly. The computations for the above currents involve the local density of states (DoS) $\nu_j(\epsilon)$ of QD $j$ at the weak link through the transport integrals:
\begin{equation}
\mathcal{L}_n(T)=\frac{1}{4T}\!\int^\infty_{-\infty}\!\!\frac{\epsilon^n}{\cosh^2\!\left(\frac{\epsilon}{2T}\right)}\mathcal{T}(T,\epsilon)d\epsilon , \,\,\,\,n=0,1,2
\label{trans_ints}
\end{equation}
where
\begin{equation}
\mathcal{T}(T,\epsilon)=2\pi e^2|t|^2\nu_1(T,\epsilon)\nu_2(T,\epsilon)
\label{trans_fact}
\end{equation}
is the transmission coefficient, and $|t|$ is the tunneling amplitude of the central link. The thermoelectric coefficients are related to the transport integrals as $G=\mathcal{L}_0$, $G_T=-\mathcal{L}_1/T$, and $G_H=\mathcal{L}_2/T$. In the  spirits of Matveev-Andreev theory \cite{andreevmatveev}, the DoS $\nu_j(T,\epsilon)$ is related to the correlation function $K_{j}\left(1/2T+it\right)$ as 
\begin{equation}
\nu_j(T,\epsilon)=\nu_{0,j}T\cosh\left(\frac{\epsilon}{2T}\right)\!\!\!\int^\infty_{-\infty}\!\!\frac{e^{i\epsilon t} K_{j}\left(\frac{1}{2T}+it\right)}{\cosh(\pi Tt)}dt,
\label{DoSdef}
\end{equation}
where $\nu_{0,j}$ stands for the DoS of the QD$_j$ which is no longer renormalized by the electron-electron interactions, while the correlation function $K_{j}\left(1/2T+it\right)$ characterizes for these interactions. The details of the derivative procedure for electric conductance and thermoelectric coefficient have been represented in Refs. \cite{thanh2018,Thanh_VN_2}. At the end, one can write the formulas of the electric conductance, thermoelectric coefficient, and thermal conductance as} \cite{Onsager}:
\begin{eqnarray}
G&=&{\color{black}\left.\frac{\partial I_e}{\partial\Delta V}\right|_{\Delta T=0}=}\frac{\pi}{2}G_{C}T\!\!\int_{-\infty}^{\infty}\!\!\frac{dt}{\cosh^{2}(\pi Tt)}\nonumber\\
&&\times K_{1}\!\left(\frac{1}{2T}+it\right)\!K_{2}\!\left(\frac{1}{2T}-it\right),\;\;\;\label{eq:cond_gen}
\end{eqnarray}
\begin{eqnarray}
&&G_{T}{=\color{black}\left.\frac{\partial I_e}{\partial\Delta T}\right|_{\Delta V=0}}= -\frac{i\pi G_{C}}{4e}\int_{-\infty}^{\infty}\frac{dt}{\cosh^{2}(\pi Tt)}\nonumber \\
 &  & \times\left\{\left[\partial_{t}K_{1}\left(\frac{1}{2T}+it\right)\right]K_{2}\left(\frac{1}{2T}-it\right)\right.\nonumber \\
 &  & -\left.K_{1}\left(\frac{1}{2T}+it\right)\left[\partial_{t}K_{2}\left(\frac{1}{2T}-it\right)\right]\right\},\label{eq:thercond_gen}
\end{eqnarray}
{\color{black} and 
\begin{eqnarray}
\mathcal{K}=\left.\frac{\partial I_h}{\partial\Delta T}\right|_{I_e=0}=G_H-T\frac{G_T^2}{G},
\label{thermal_cond}
\end{eqnarray}
with $G_H$ is thermal coefficient which is expressed as
\begin{eqnarray}
&&G_{H}=\left.\frac{\partial I_h}{\partial\Delta T}\right|_{\Delta V=0}=\frac{\pi G_{C}}{2e^{2}}\int_{-\infty}^{\infty}dt\nonumber\\
&&\times\left\{\frac{\pi^{2}T^{2}\left[2-\cosh^{2}(\pi Tt)\right]}{\cosh^{4}(\pi Tt)}\! K_{1}\!\left({\displaystyle\!\frac{1}{2T}+it}\!\right)\!K_{2}\!\left({\displaystyle\!\frac{1}{2T}-it}\!\right)\right.\nonumber \\
& &
\left.+\frac{1}{\cosh^{2}(\pi Tt)}\partial_{t}K_{1}\left({\displaystyle \frac{1}{2T}+it}\right)\partial_{t}K_{2}\left({\displaystyle \frac{1}{2T}-it}\right)\right\},
\label{eq:ther_coe_gen}
\end{eqnarray}
where $G_{C}=2\pi e^2 \nu_{0,1}\nu_{0,2}|t|^2$ is a conductance of the central (tunnel) area assuming that the electrons in the QDs are noninteracting.} The TP in the linear regime is defined at $I_{\color{black} e}=0$ as 
{\color{black}
\begin{equation}
S=-\left.\frac{\Delta V}{\Delta T}\right|_{I_e=0}=\frac{G_{T}}{G}.
\end{equation}
The definition of the figure of merit $ZT$ which characterizes the quality of a thermoelectric material, reads 
\begin{equation}
ZT=\frac{GTS^2}{\mathcal{K}}.
\label{ZTeq}
\end{equation}
One should notice that the equation for the thermal conductance $\mathcal{K}$ (\ref{thermal_cond}) contains both the diagonal and the off-diagonal  Onsager coefficients \cite{Onsager}. The thermal conductance is connected to the electric conductance $G$ (typically, in the low-temperature regime) by a universal constant called the Lorenz number \cite{Zlatic_book,Benenti}:
\begin{equation}
\frac{\mathcal{K}}{GT}=L_0=\frac{\pi^2}{3e^2}.
\label{WF}
\end{equation}
Connection between electric and thermal conductances (\ref{WF}) is established by the WF law.  The validity of WF law is attributed to the fact that both charge and heat are transferred by the same quasiparticles. 
In some strongly correlated systems,  however, the $\mathcal{K}/(GT)$ is deviated from $L_0$ still remaining the universal number \cite{Karki_2020,Kis2023}.   
Thus, the generalized WF law is applied for description of such systems.} The computation of thermoelectric coefficients in Eqs. (\ref{eq:cond_gen}), (\ref{eq:thercond_gen}), and (\ref{eq:ther_coe_gen}) \cite{misprint}
requires the explicit form of the correlation functions $K_{1,2}\left(1/2T\pm it\right)$.

\section{Correlation function $K_{j}(\tau)$: \label{sec:Correlationfunction}}

{\color{black}In the Matveev-Andreev spirits, the time-ordered correlation function $K_{j}(\tau)=\langle T_\tau F(\tau)F^\dagger(0)\rangle$ ($T_\tau$ is the time-ordering operator, the imaginary time t runs from $0$ to $\beta=1/T$) accounts for interaction effects in QDs. The operator $F^\dagger(0)$ increases number of electrons entering the $\text{QD}_{j}$ through the weak link (characterized by operator $\hat{n}$) from $0$ to $1$ at time $t=0$ and $F(\tau)$ decreases it back to $0$ at time $t=\tau$,} one can replace $\hat{n}$ by $n_{j\tau}\left(t\right)=\theta\left(t\right)\theta\left(\tau-t\right)$ with $\theta\left(t\right)$ is the unit step function. Therefore, the correlation function $K_{j}(\tau)$ is computed through the functional integration over the bosonic fields $K_{j}(\tau)=Z_{j}(\tau)/Z_{j}(0)$ \cite{andreevmatveev}.

\subsection{The 1CK case: Perturbative solution:}

In the case one CKC is settled down in the FL-1CK state by decoupling
one of the two QPCs, the functional integral writes 
\begin{eqnarray}
\!\!\!\!\!\!\!\!Z_{j}(\tau)\! & = & \!\!\int\!\mathcal{D}\phi_{j}\exp\left[-\mathcal{S}_{0,j}-\mathcal{S}_{C,j}(\tau)-\mathcal{S}_{s,j}\right],\label{func_intergal_1CK}
\end{eqnarray}
where $\mathcal{S}_{0,j}$, $\mathcal{S}_{C,j}$, and $\mathcal{S}_{s,j}$
are Euclidean actions describing the free (non-interacting) one-dimensional Fermi gas, Coulomb blockade
in the QD and the backscattering at the QPC of the CKC $j$, respectively.
They are written as \cite{andreevmatveev,Aleiner98,giamarchi}
\begin{eqnarray}
\!\!\!\!\mathcal{S}_{0,j}\!\! & = & \!\frac{v_{F}}{2\pi}\int_{0}^{\beta}\!\!\!\!dt\!\!\int\!\!dx\!\left[\frac{(\partial_{t}\phi_{j})^{2}}{v_{F}^{2}}+(\partial_{x}\phi_{j})^{2}\right],\label{freeaction_1ck}\\
\!\!\!\!\mathcal{S}_{C,j}\!\! & = & \!E_{C,j}\!\int_{0}^{\beta}\!\!\!\!dt\left[n_{j\tau}(t)+\frac{1}{\pi}\phi_{j}(0,t)-N_{j}\right]^{2}\!,\label{charge_action_1CK}\\
\mathcal{S}_{s,j}\!\! & = & \!\!-\frac{2D}{\pi}|r_{j}|\!\int_{0}^{\beta}\!\!\!dt\cos\left[2\phi_{j}(0,t)\right].\label{BS_1CK}
\end{eqnarray}
with {\color{black} $\phi_{j}$ represents the bosonic field at the QPC of the CKC $j$, and $v_{F}$ is the
Fermi velocity, $E_{C,j}$ is the charging energy of the QD $j$,  $N_{j}$ is the normalized dimensionless gate voltage, controlled
by plunger gates (not shown in Fig. \ref{Fig1}), and $D$ is a bandwidth}. One should notice that the bosonic field describing
the electrons moving through the constriction is blocked by the Coulomb
interaction in the QD. Therefore, $Z_{j}(\tau)$ can be computed perturbatively
over $|r_{j}|$ for the small backscattering at the QPC ($|r_{j}|\ll1$),
and the correlation function $K_{j}(\tau)$ then is 
\begin{eqnarray}
K_{j}\!\left(\tau\right) & = & \!\left(\frac{\pi^{2}T}{\gamma E_{C,j}}\right)^{2}\!\!\!\frac{1}{\sin^{2}\left(\pi T\tau\right)}\left[1-2\gamma\xi|r_{j}|\cos\left(2\pi N_{j}\right)\right.\nonumber \\
 &  & \left.+4\pi^{2}\xi\gamma|r_{j}|\frac{T}{E_{C,j}}\sin\left(2\pi N_{j}\right)\cot\left(\pi T\tau\right)\right]~,\label{KFL}
\end{eqnarray}
with $\gamma=e^{C}$, $C\approx0.577$ is Euler's constant, $\xi=1.59$
is a numerical constant \cite{andreevmatveev}. 

\subsection{The symmetric 2CK case: Nonperturbative solution:}

For convenient calculation later, one can define the variables $\phi_{j,\rho/\sigma}=\phi_{j,1}\pm\phi_{j,2}$
so-called charge/spin fields {\color{black} where $\phi_{j,\alpha}$ ($\alpha=1,2$) represents the bosonic field at the QPC $\alpha$ of the CKC $j$}. The functional integral in this case is written as: 
\begin{eqnarray}
\!\!\!\!\!\!\!\!Z_{j}(\tau)\! & = & \!\!\!\!\!\prod_{\lambda=\rho,\sigma}\!\int\!\mathcal{D}\phi_{j,\lambda}\exp\left[-\mathcal{S}_{0,j}-\mathcal{S}_{C,j}(\tau)-\mathcal{S}_{s,j}\right],\label{func_int_2CK}
\end{eqnarray}
where $\mathcal{S}_{0,j}$, $\mathcal{S}_{C,j}$, and $\mathcal{S}_{s,j}$
are Euclidean actions describing the free Fermi liquid, Coulomb blockade
in the QD and the backscattering at the QPCs of the CKC $j$,
respectively. The action $\mathcal{S}_{0,j}$ is presented as a sum
of two independent actions 
\begin{eqnarray}
\!\!\!\!\mathcal{S}_{0,j}\!\! & = & \!\!\!\!\sum_{\lambda=\rho,\sigma}\!\frac{v_{F}}{2\pi}\int_{0}^{\beta}\!\!\!\!dt\!\!\int\!\!dx\!\left[\frac{(\partial_{t}\phi_{j,\lambda})^{2}}{v_{F}^{2}}+(\partial_{x}\phi_{j,\lambda})^{2}\right].\label{free_action_2CK}
\end{eqnarray}
The Coulomb blockade action $\mathcal{S}_{C,j}$ in bosonic representation
reads \cite{Aleiner98}
\begin{eqnarray}
\!\!\!\!\mathcal{S}_{C,j}\!\! & = & \!E_{C,j}\!\int_{0}^{\beta}\!\!\!\!dt\left[n_{j\tau}(t)+\frac{\sqrt{2}}{\pi}\phi_{j,\rho}(0,t)-N_{j}\right]^{2}\!.\label{charge_action_2CK}
\end{eqnarray}
The contribution $\mathcal{S}_{s,j}$ in the action of each CKC characterizes
the weak backscattering at the QPCs is 
\begin{eqnarray}
\!\!\mathcal{S}_{s,j}\!\!=\!\!-\frac{2D}{\pi}|r_{j}|\!\int_{0}^{\beta}\!\!\!dt\cos\left[\sqrt{2}\phi_{j,\rho}(0,t)\right]\cos\left[\sqrt{2}\phi_{j,\sigma}(0,t)\right].\;\;\;\;\label{BS_2CK}
\end{eqnarray}

In the absence of backscattering $|r_{j}|=0$, the functional integral
Eq.(\ref{func_int_2CK}) is Gaussian. The correlator $K_{j}^{(0)}(\tau)\equiv K_{j}(\tau)|_{r=0}=K_{j,\rho}(\tau)$
is computed at low temperature $T\ll E_{C}$ and at $\tau\gg E_{C}^{-1}$:
\begin{equation}
K_{j,\rho}(\tau)=\frac{\pi^{2}T}{2\gamma E_{C,j}}\frac{1}{|\sin(\pi T\tau)|}.\label{corr}
\end{equation}
The perturbative results (see Ref.\cite{andreevmatveev})
showed that the thermoelectric properties of the system are controlled
by charge and spin fluctuations at low frequencies (below $E_{C,j}$).
One should notice that the effect of small but finite $|r_{j}|$ on
the charge modes is negligible in comparison with the Coulomb blockade
but it changes the low frequency dynamics of the unblocked spin modes
dramatically. The correlation function can be split into charge and
spin components as $K_{j}(\tau)=K_{j,\rho}(\tau)K_{j,\sigma}(\tau)$,
with $K_{j,\sigma}(\tau)=Z_{j,\sigma}(\tau)/Z_{j,\sigma}(0)$. We simply
replace the $\cos\left[\sqrt{2}\phi_{j,\rho}(0,t)\right]$ in action
Eq. (\ref{BS_2CK}) by the $\langle\cos[\sqrt{2}\phi_{j,\rho}(0,t)]\rangle_{\tau}=\sqrt{2\gamma E_{C,j}/\pi D}\cos\left[\pi N_{j}-\chi_{j\tau}(t)\right]$,
with $\chi_{j}(t)=\pi n_{j\tau}(t)+\delta\chi_{j\tau}(t)$ , $\delta\chi_{j\tau}(t)\approx(\pi^{2}T/2E_{C,j})\left[\cot(\pi T(t-\tau)-\cot(\pi Tt)\right]$ and
obtain the effective action for the spin degrees of freedom in the
form 
\begin{eqnarray}
\!\!\!\!\mathcal{S}_{\tau j}\! & = & \!\!\int\!\!dx\int_{0}^{\beta}\!\!\!\!dt\frac{v_{F}}{2\pi}\left[\frac{(\partial_{t}\phi_{j,\sigma})^{2}}{v_{F}^{2}}+(\partial_{x}\phi_{j,\sigma})^{2}\right]\nonumber \\
 &  & -\int_{0}^{\beta}dt\sqrt{\frac{4D}{v_{F}}}\tilde{\lambda}_{j\tau}(t)\cos\left[\sqrt{2}\phi_{j,\sigma}(0,t)\right],\label{eq:S_spin}
\end{eqnarray}
where 
\begin{eqnarray}
 &  & \tilde{\lambda}_{j\tau}(t)=\Lambda_{j}(-1)^{n_{\tau}(t)}\cos\left[\pi N_{j}-\delta\chi_{j\tau}(t)\right],\nonumber \\
 &  & \Lambda_{j}=|r_{j}|\sqrt{\frac{2\gamma v_{F}E_{C,j}}{\pi D}}.
\end{eqnarray}

After performing the refermionization, our model [as shown in Eq.
(\ref{eq:S_spin})] is mapped onto an effective Anderson model,
which is described by Hamiltonian 
\begin{equation}
H_{j,\tau}^{\rm eff}\!\left(t\right)\!=\!\!\int\!\left[v_{F}kc_{j,k}^{\dagger}c_{j,k}\!-\tilde{\lambda}_{j\tau}\!(t)(c+c^{\dagger})\!\left(c_{j,k}\!-c_{j,k}^{\dagger}\right)\!\right]\!dk,\label{eq:Heff}
\end{equation}
in which the operators $c_{j,k}^{\dagger}$ and $c_{j,k}$ satisfying
the anti-commutation relations$\left\{ c_{j,k},c_{j,k^{'}}^{\dagger}\right\} =\delta\left(k-k^{'}\right)$
create and destroy chiral fermions; $c$ is a local fermionic annihilation operator anti-commuting with $c_{j,k}^{\dagger}$ and $c_{j,k}$. We see that the model is free and equivalent to a resonant level model
where the leads are coupled to Majorana fermion $\eta=(c+c^{\dagger})/\sqrt{2}$
on the impurity. The time dependent Hamiltonian (\ref{eq:Heff}) can
be split into $H_{j,0}^{\rm eff}+H_{j,\tau}^{\prime}\left(t\right)$ by
replacing $\tilde{\lambda}_{j\tau}(t)\rightarrow\tilde{\lambda}_{j\tau}(t)/\left(-1\right)^{n_{\tau}\left(t\right)}$.
The time-independent Hamiltonian part $H_{j,0}^{\rm eff}$ is $H_{j,\tau=0}^{\rm eff}$
while the correction is 
\begin{eqnarray}
H_{j,\tau}^{\prime}\left(t\right)=2\Lambda_{j}\!\left\{ \cos\left[\pi N_{j}\right]-\cos\left[\pi N_{j}\!-\delta\chi_{j\tau}(t)\right]\right\} \eta\zeta,\;\;\;\;\;\;
\end{eqnarray}
with $\zeta=\int_{-\infty}^{\infty}\left(c_{j,k}-c_{j,k}^{\dagger}\right)dk/\sqrt{2}$
describes the Majorana fermion of the leads in the resonant level
model. Our solution, being nonperturbative in $|r_{j}|$ and accounting for low-frequency dynamics of the spin modes, leads to the appearance of the Kondo-resonance width $\Gamma_{j}$ in the vicinity of Coulomb peaks 
\begin{eqnarray}
\Gamma_{j}\left(N_{j}\right)=\frac{8\gamma E_{C,j}}{\pi^{2}}|r_{j}|^{2}\cos^{2}(\pi N_{j}).\label{eq:Gam}
\end{eqnarray}
We then compute the correlation function straightforwardly and obtain the zero-order in $|r_j|$ term corresponding to the Hamiltonian part $H_{j,0}^{\rm eff}$ as
\begin{eqnarray}
 &  & K_{j}^{(0)}\!\left(\frac{1}{2T}+it\right)=\frac{\pi T\Gamma_{j}}{\gamma E_{C,j}}\frac{1}{\cosh(\pi Tt)}\nonumber \\
 &  & \times\int_{-\infty}^{\infty}\frac{e^{\omega\left(1/2T+it\right)}}{\left(\omega^{2}+\Gamma_{j}^{2}\right)\left(1+e^{\omega/T}\right)}d\omega,\label{eq:K0}
\end{eqnarray}
and the first-order term when the correction Hamiltonian part $H_{j,\tau}^{\prime}\left(t\right)$
is taken into account, is 
\begin{eqnarray}
 &  & K_{j}^{(1)}\left(\frac{1}{2T}+it\right)=\!-\frac{4T}{E_{C,j}}\frac{|r_{j}|^{2}\sin\left(2\pi N_{j}\right)}{\cosh(\pi Tt)}\nonumber \\
 &  & \times\ln\left(\frac{E_{C,j}}{T+\Gamma_{j}}\right)\int_{-\infty}^{\infty}\!\!\!\!\!\!d\omega\frac{\omega e^{\omega\left(1/2T+it\right)}}{\left(\omega^{2}+\Gamma_{j}^{2}\right)\left(1+e^{\nicefrac{\omega}{T}}\right)}~.\label{eq:K1}
\end{eqnarray}
\begin{figure}
\includegraphics[width=.55\textwidth,left]{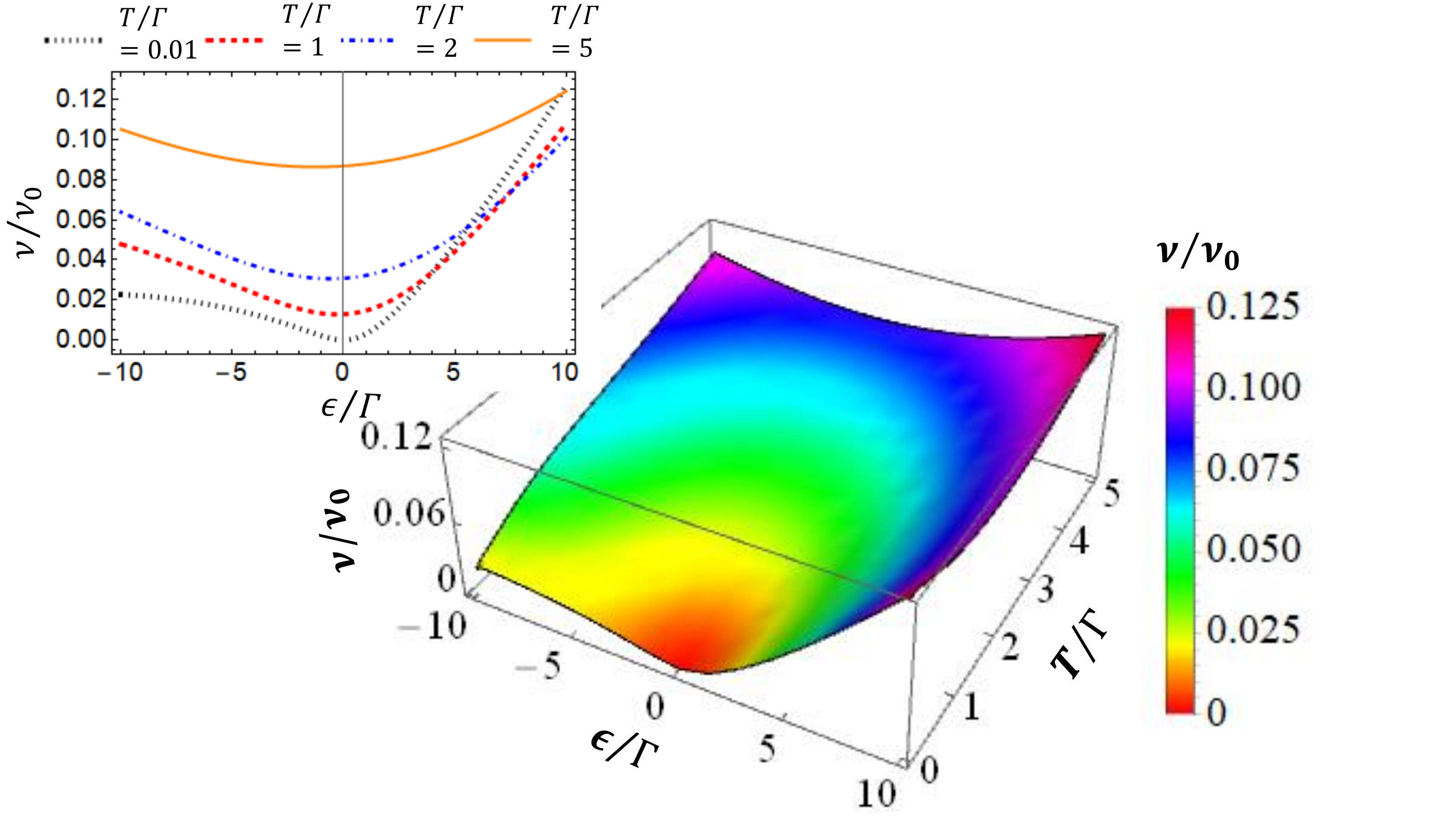}
\vspace{-0.5cm}
\caption{The density of states (DoS) of a charge Kondo circuit $\nu/\nu_{0}$ is plotted. Both the even function of time correlation functions $K_{j}^{(0)}$ given by the Eq. (\ref{eq:K0}) and the odd function of time $K_{j}^{(1)}$ given by Eq. (\ref{eq:K1}) being proportional to the second power of the particle-hole symmetry breaking parameter $|r|$ are taken into account. The DoS is plotted as a function of ratio of energy to the Kondo resonance width $\epsilon/\Gamma$ and ratio of temperature to the Kondo resonance width $T/\Gamma$ (we choose $N_1=N_2=N=0.42$, $|r_1|^2=|r_2|^2=|r|^2=0.12$, $E_{C,1}=E_{C,2}=E_C$, and therefore $\Gamma_1=\Gamma_2=\Gamma$). Insert: zoomed in DoS  as the function of ratio of energy to the Kondo resonance width $\epsilon/\Gamma$ at different temperatures $T/\Gamma=0.01,\,1,\,2,\,5$ corresponding to the black dotted, red dashed, blue dot-dashed, and orange lines.}
\label{DoS}
\end{figure}
One should notice that if the correlation function $K_{j}(\tau)$ is considered at the zeroth-$|r_j|$ order $K_{j}^{(0)}$ in the Eq. (\ref{eq:K0}), the DoS of a charge Kondo circuit $j$ is an even function of the energy $\epsilon/\Gamma_j$, while the contribution to DoS due to the first order in $|r_j|$ term $K_{j}^{(1)}$ in the Eq. (\ref{eq:K1}) is an odd function of energy. The DoS of a CKC is shown in Fig. \ref{DoS}. In the regime of small compared to the temperature  energies (in comparison to the Kondo temperature which is $E_{C,j}$ in the charge Kondo model) the DoS are analytic functions obeying a Taylor expansion $\nu_{j}/\nu_{0,j} \propto a_j(|r_j|,T)+b_j(|r_j|,T)\left(\epsilon/E_{C,j}\right)+c_j(|r_j|,T)\left(\epsilon/E_{C,j}\right)^2+d_j(|r_j|,T)\left(\epsilon/E_{C,j}\right)^3 + ...$ (here $a_j,b_j,c_j,d_j$ are constant depending on the system parameters) [see the insert of Fig.\ref{DoS})] while $\nu_{j}/\nu_{0,j} \propto \left|\epsilon/E_{C,j}\right|$ in the other regime of energy. The latter is explained based on the orthogonality catastrophe \cite{furusakimatveev,Kis2023}. The symmetry and asymmetry of the contributions originated from $K^{(0)}_j$ and $K^{(1)}_j$ to the DoS implies that it is enough to take into account $K^{(0)}_j$ in order to compute $G$ and $G_H$ but it is necessary to consider $K^{(1)}_j$ when one calculates $G_T$ [see Eqs. (\ref{trans_ints},\ref{trans_fact})].
The formulas (\ref{eq:K0}) and (\ref{eq:K1}) will be used to calculate the thermoelectric coefficients in the next Section.

\section{Main results}

{\color{black} Going beyond the perturbation theory assuming smallness of the reflection amplitudes of the QPCs as represented in Ref. \cite{thanh2018}, we develop a controllable and reliable approach for the quantitative description of the Fermi-to-non-Fermi liquid crossovers and interplay around the intermediate
coupling fixed points in this work.}

\subsection{Weak coupling between single- and two-channel charge Kondo circuits}

{\color{black} To begin with description of the weakly coupled single and two channel Kondo simulators} we consider the circuit consisting of  the left CKC being in the FL-1CK
state and the right CKC operating in the NFL-2CK state. We apply the correlation
functions $K_{1}\left(\frac{1}{2T}+it\right)$ and $K_{2}\left(\frac{1}{2T}-it\right)$
as shown in Eqs. (\ref{KFL}) and (\ref{eq:K0}-\ref{eq:K1}), respectively.
The electric conductance is {\color{black} given by}
\begin{equation}
G=\frac{\pi^{2}G_{C}T^{3}}{96\gamma^{3}E_{C,1}^{2}E_{C,2}}F_G\left(\frac{\Gamma_{2}}{T}\right),\label{eq:G_1CK_2CK}
\end{equation}
with $F_G$ is a dimensionless {\color{black} function describing} the {\color{black} interplay} between the {\color{black} width of the} Kondo resonance of the right CKC $\Gamma_{2}$ and the temperature $T$. It is {\color{black} expressed} as
\begin{align}
F_G\left(p_{2}={\color{black} \frac{\Gamma_2}{T}}\right) & =\!\int_{-\infty}^{\infty}\!\!\!du\,J\left(p_{2},u\right),
\end{align}

\begin{align}
\,J\left(p_{2},u\right) & =\frac{{\color{black}p_2}\left[u^{2}+\pi^{2}\right]\left[u^{2}+9\pi^{2}\right]}{\cosh^{2}\left(\frac{u}{2}\right)\left[u^{2}+p_{2}^{2}\right]}.
\end{align}
The thermoelectric coefficient {\color{black} $G_T$} is given by
\begin{eqnarray}
&& G_{T}=-\frac{\pi^{5}\xi G_C}{72e\gamma^2}|r_{1}|\sin\left(2\pi N_{1}\right)\frac{{\color{black}T^4}}{E_{C,1}^3E_{C,2}}F_G\left(\frac{\Gamma_{2}}{T}\right)\nonumber \\
&& -\frac{\pi^3\xi G_C}{180e\gamma^2}|r_{1}|\sin\left(2\pi N_1\right)\frac{{\color{black}T^4}}{E_{C,1}^3E_{C,2}}F_T\left(\frac{\Gamma_2}{T}\right)\nonumber \\
&& -\frac{\pi G_C}{{\color{black}40}e\gamma^2}|r_2|^2\sin\left(2\pi N_2\right)\nonumber\\
&&\times\frac{{\color{black}T^4}}{E_{C,1}^2E_{C,2}{\color{black}\Gamma_2}}\ln\left(\frac{E_{C,2}}{T+\Gamma_2}\right)F_T\left(\frac{\Gamma_2}{T}\right),\label{eq:GT_1CK_2CK}
\end{eqnarray}
with 
\begin{align}
F_{T}\left(p_{2}\right) & =\!\int_{-\infty}^{\infty}\!\!\!du\,u^{2}J\left(p_{2},u\right).
\end{align}
Following the discussion in Ref. \cite{thanh2018}, based on the perturbative
solution, the Seebeck effect on a weak link between 1CK and 2CK is
characterized by the {\color{black} interplay} between the Fermi and non-Fermi liquid {\color{black} regimes}
(see Eq. (24) in the Ref. \cite{thanh2018}). However, {\color{black} this 
competition is appreciable only at sufficiently
high temperatures $T\gg\Gamma_{2}$. }At the very low temperatures,
$T\ll\Gamma_{2}$, the {\color{black} Fermi liquid regime characterized by the linear dependence of the TP as the function the temperature holds.} 

\subsubsection{$T\gg\Gamma_{2}$ limit: Fermi-liquid on the left and non Fermi-liquid on the right CKC:}
At temperature {\color{black} regime}  $T\gg\Gamma_{2}${\color{black}, we get  $F_G(p_2= \Gamma_2/T\rightarrow 0)=9\pi^5$ and $F_T(p_2\rightarrow 0)=256\pi^4p_2/5$. Naturally,} the expression in Eq. (\ref{eq:GT_1CK_2CK}) reproduces the perturbative result {\color{black} (see Eq. (23) of Ref.\cite{thanh2018})}. The
{\color{black} equation for TP} is thus similar to the formula (24) in Ref.\cite{thanh2018}:
\begin{eqnarray}
S & = & -\frac{4\pi^{3}\xi\gamma}{3e}|r_{1}|\frac{T}{E_{C,1}}\sin\left(2\pi N_{1}\right)\nonumber \\
 &  & -\frac{{\color{black}256}\gamma}{{\color{black}75}\pi^{2}e}|r_{2}|^{2}\ln\left(\frac{E_{C,2}}{T}\right)\sin\left(2\pi N_{2}\right).
\label{mixed}
\end{eqnarray}
The crossover  {\color{black} temperature $\widetilde{T}$}  separating two contributions ({\color{black} T-linear FL and log T NFL}) in the TP is defined
as follows:
{\color{black}
\begin{equation}
\ln\left(\frac{E_{C,2}}{\widetilde{T}}\right)\frac{E_{C,1}}{\widetilde{T}}=\frac{25\pi^5\xi}{256}\frac{|r_{1}|}{|r_{2}|^{2}}.
\end{equation}
}
If $\Gamma_2\ll T\ll \widetilde{T}$, NFL-2CK behavior of the TP is predicted to be {\color{black} dominant}. In the opposite limit, $T\gg \widetilde{T}\gg \Gamma_2$, the FL-1CK regime with the weak NFL-2CK corrections is expected.

\subsubsection{$T\ll\Gamma_{2}$ limit:  Fermi-liquid regime}
At temperature {\color{black} regime} $T\ll\Gamma_{2}${\color{black}, we get $F_G(p_2=\Gamma_2/T\gg 1)=256\pi^4/5p_2$ and $F_T(p_2\gg 1)=256\pi^6/7p_2$}. The expression in Eq. (\ref{eq:GT_1CK_2CK}) {\color{black} produces} the linear in temperature term. 
The {\color{black} FL picture is thus described through the property of TP as}
{\color{black}
\begin{eqnarray}
&& S=-\frac{12\pi^{3}\xi\gamma}{7e}\left[\frac{|r_{1}|}{E_{C,1}}\sin\left(2\pi N_{1}\right)\right.\nonumber\\
&& \left.\!\!+\frac{\tan\left(\pi N_{2}\right)}{4\gamma\xi E_{C,2}}|r_{2}|^{2}\ln\!\left(\!\frac{E_{C,2}}{\Gamma_{2}}\!\right)\!\right]T.
\label{tanh}
\end{eqnarray}
}
{\color{black} In summary, in the regime of the weak coupling between single- and two-channel charge Kondo circuits, there exist two energy scales: the Kondo width $\Gamma_2\sim(8\gamma/\pi^2)E_{C,2}|r_2|^2$ and  the crossover temperature $\widetilde{T}$.  When the condition 
$0$$<$$ \Gamma_2$$<$$\tilde T$$<$$E_{C,2}$ holds,  the domain of validity for the Fermi liquid regime is significantly 
larger compared to the Non-Fermi liquid domain. The results discussed in this Section are completely consistent with the perturbative results represented in Ref. \cite{thanh2018}.  The mixed FL+NFL operational regime of the thermopower 
given by Eq.  (\ref{mixed}) can only be achieved  at the intermediate temperatures restricted by the condition  $(8\gamma/\pi^2)E_{C,1}|r_1|^2\ll T \ll \left( E_{C,1}, E_{C,2}\right)$.}

\subsection{Weak coupling between {\color{black} two-site} two-channel charge Kondo circuits} \label{2CK_2CK}

{\color{black} To compute electric conductance through weakly linked  two-site two-channel charge Kondo circuit it is sufficient to take into account only even in time part of the kernels $K_{j}^{(0)}\left(\frac{1}{2T}+it\right)$.  As a result, we get}
\begin{equation}
G=\frac{G_{C}T^{2}}{24\gamma^{2}E_{C,1}E_{C,2}}F_{C}\left(\frac{\Gamma_{1}}{T},\frac{\Gamma_{2}}{T}\right),\label{eq:genG}
\end{equation}
{\color{black} where}
\begin{equation}
F_{C}\left(p_{1},p_{2}\right)=\int_{-\infty}^{\infty}dz\int_{-\infty}^{\infty}du\,F\left(p_{1},p_{2},z,u\right),\label{FC}
\end{equation}
\begin{eqnarray}
&&F\left(p_{1},p_{2},z,u\right)=\frac{p_{1}p_{2}u\left[u^{2}+4\pi^{2}\right]}{\sinh\left(\frac{u}{2}\right)\left[\cosh\left(z\right)+\cosh\left(\frac{u}{2}\right)\right]}\nonumber \\
&&\times\frac{1}{\left[\left(z+\frac{u}{2}\right)^{2}+p_{1}^{2}\right]\left[\left(z-\frac{u}{2}\right)^{2}+p_{2}^{2}\right]}.\label{FFF}
\end{eqnarray}
The integral in Eq. (\ref{eq:thercond_gen}) {\color{black} vanishes due to the particle-hole (PH) symmetry when even in time contribution of both kernels $K_{1,2}^{(0)}\left(\frac{1}{2T}\pm it\right)$ is taken into account. As a result the off-diagonal coefficient $G_{T}^{(0)}=0$.} We therefore need to consider the first non-vanishing {\color{black} correction with respect to the PH-symmetry breaking parameter. By taking into account even-odd products of the kernels}
$K_{1}^{(0)}\left(\frac{1}{2T}+it\right)K_{2}^{(1)}\left(\frac{1}{2T}-it\right)$
and $K_{1}^{(1)}\left(\frac{1}{2T}+it\right)K_{2}^{(0)}\left(\frac{1}{2T}-it\right)$,
we obtain the lowest order (we consider the model in the vicinity of the intermediate coupling fixed point) non-zero contribution to thermoelectric coefficient as follows
\begin{eqnarray}
\!\!\!\! &  & \!\!\!\!G_{T}=-\frac{G_{C}T^{3}}{{\color{black}6} e\gamma\pi E_{C,1}E_{C,2}}\nonumber \\
\!\!\!\! &  & \!\!\!\!\!\!\!\!\times\left\{ \frac{|r_{1}|^{2}}{\Gamma_{1}}\ln\left(\frac{E_{C,1}}{T+\Gamma_{1}}\right)\sin\left(2\pi N_{1}\right)F_{T,s}\left(\frac{\Gamma_{1}}{T},\frac{\Gamma_{2}}{T}\right)\right.\nonumber \\
\!\!\!\! &  & \!\!\!\!\!\!\!\!\!\!\!+\left.\frac{|r_{2}|^{2}}{\Gamma_{2}}\ln\left(\frac{E_{C,2}}{T+\Gamma_{2}}\right)\sin\left(2\pi N_{2}\right)F_{T,m}\left(\frac{\Gamma_{1}}{T},\frac{\Gamma_{2}}{T}\right)\right\} ,\label{eq:genGT}
\end{eqnarray}
where 
\begin{equation}
\!\!\!\!F_{T,s}\!\left(p_{1},p_{2}\right)=\!\!\int_{-\infty}^{\infty}\!\!\!\!\!\!dz\!\!\int_{-\infty}^{\infty}\!\!\!\!\!\!du\left(z+\frac{u}{2}\right)zF\left(p_{1},p_{2},z,u\right),\label{FTC1}
\end{equation}
\begin{align}
\!\!\!\!F_{T,m}\!\left(p_{1},p_{2}\right)=\!\!\!\int_{-\infty}^{\infty}\!\!\!\!\!\!dz\!\!\int_{-\infty}^{\infty}\!\!\!\!\!\!du\left(z-\frac{u}{2}\right)zF\left(p_{1},p_{2},z,u\right).\label{FTC2}
\end{align}
The Eqs. (\ref{eq:genG}-\ref{eq:genGT}) {\color{black} represent} the central results of this part. By varying parameters such as temperature, gate voltages, and/or reflection amplitudes at the QPCs, one can achieve four different regimes of the thermoelectric transport. The details of the calculations for the electric conductance and the thermal coefficient are represented in the Appendix. We {\color{black} derive and analyse the asymptotic equations} for the TP in each regime in four segments below. 

\subsubsection{$\left(\Gamma_{1},\:\Gamma_{2}\right)\ll T$, non-Fermi-liquid regime:} \label{case1}

The TP demonstrates the weak NFL behavior at ``high'' temperature: $T\gg\left(\Gamma_{1},\:\Gamma_{2}\right)$ as 
\begin{eqnarray}
S & = & -{\color{black}\frac{9\gamma}{8e}}\left[|r_{1}|^{2}\ln\left(\frac{E_{C,1}}{T}\right)\sin\left(2\pi N_{1}\right)\right.\nonumber \\
 &  & \left.+|r_{2}|^{2}\ln\left(\frac{E_{C,2}}{T}\right)\sin\left(2\pi N_{2}\right)\right].\label{eq:S_2NFL}
\end{eqnarray}
The similarity between Eq. (\ref{eq:S_2NFL}) and Eq. (28) of Ref. \cite{thanh2018} implies that the regime $T\gg\left(\Gamma_{1},\:\Gamma_{2}\right)$ {\color{black} mimics} the perturbative result. The Kondo-resonance {\color{black} width} $\Gamma_{j}$ {\color{black} vanishes} at the Coulomb peaks and increased when the gate voltage $N_{j}$ goes out of the half integer values. This {\color{black} regime is accessible} at the centre of the $(N_1, N_2)$ window (if one considers $0\le N_1,N_2 \le 1$). Due to the logarithmic dependent on temperature but small value of TP [see Eq. (\ref{eq:S_2NFL})] {\color{black} we refer to it as} a weak NFL {\color{black} scenario}.

\subsubsection{$\Gamma_{1}\ll T\ll\Gamma_{2}$, {\color{black} non-Fermi-liquid} on the left and Fermi-liquid on the right CKC:}\label{case2}

Let us recall that the Kondo resonances' widths  $\Gamma_{1},\: \Gamma_{2}$ depend on the gate voltages and therefore {\color{black} represent the energy scales competing} with temperature effects  in the vicinity of the Coulomb peaks. {\color{black} At the parametric regime $\Gamma_{1}\ll T\ll\Gamma_{2}$ it corresponds to the domain of the gate voltages such that the QD $1$ is fine-tuned to a Coulomb peak closer than the QD $2$ ($N_1$ is closer to a half integer value)}. The TP {\color{black} consists of} two components: {\color{black} weak FL and NFL} characteristics as
\begin{eqnarray}
& & \!\!\!S=-\frac{{\color{black}1024}\gamma}{75e\pi^2}\left[|r_1|^2\ln\left(\frac{E_{C,1}}{T}\right)\sin\left(2\pi N_1\right)\right.\nonumber \\
& &\!\!\!\!\!\!\!\!\! \left.+\frac{25\pi^3}{{\color{black}256}}|r_2|^2\frac{T}{\Gamma_2}\ln\left(\frac{E_{C,2}}{\Gamma_2}\right)\sin\left(2\pi N_2\right)\right].
\end{eqnarray}
The crossover {\color{black} temperature $T^\ast$ separating} two regimes is defined as 
\begin{eqnarray}
\frac{E_{C,2}}{T^{\ast}}\ln\left(\frac{E_{C,1}}{T^{\ast}}\right) & = & \frac{25\pi^5}{{\color{black}2048}\gamma\cos^{2}\left(\pi N_2\right)}\frac{1}{|r_1|^2}\nonumber \\
 &  & \times\ln\left(\frac{\pi^2}{8\gamma|r_2|^2\cos^2\left(\pi N_2\right)}\right).
\end{eqnarray}
The NFL behavior {\color{black} dominates} if $T\ll T^\ast$ while the FL {\color{black} regime}
is predicted at the opposite limit $T\gg T^\ast$.

\subsubsection{$\Gamma_{2}\ll T\ll\Gamma_{1}$, Fermi-liquid on the left and {\color{black} non-Fermi-liquid} on the right CKC:} \label{case3}
This {\color{black} parametric regime is complementary} to the case discussed in the Section \ref{case2}. The regime is achieved
when the {\color{black} gate voltage applied to the QD $2$ tunes it closer to the Coulomb blockade peak} than the QD $1$ is. The TP is characterized by the {\color{black}weak FL on the left and NFL effect} on the right CKC as
\begin{eqnarray}
& & \!\!\!S=-\frac{{\color{black}1024}\gamma}{75e\pi^2}\left[|r_2|^2\ln\left(\frac{E_{C,2}}{T}\right)\sin\left(2\pi N_2\right)\right.\nonumber \\
& &\!\!\!\!\!\!\!\!\! \left.+\frac{25\pi^3}{{\color{black}256}}|r_1|^2\frac{T}{\Gamma_1}\ln\left(\frac{E_{C,1}}{\Gamma_1}\right)\sin\left(2\pi N_1\right)\right].
\end{eqnarray}
The crossover {\color{black} temperature $T^{\ast\ast}$ separating FL and NFL regimes} is defined as 
\begin{eqnarray}
\frac{E_{C,1}}{T^{\ast\ast}}\ln\left(\frac{E_{C,2}}{T^{\ast\ast}}\right) & = & \frac{25\pi^5}{{\color{black}2048}\gamma\cos^2\left(\pi N_1\right)}\frac{1}{|r_2|^2}\nonumber \\
 &  & \times\ln\left(\frac{\pi^2}{8\gamma|r_1|^2\cos^2\left(\pi N_1\right)}\right).
\end{eqnarray}
The NFL behavior {\color{black} dominates when} $T\ll T^{\ast\ast}$ while the FL {\color{black} regime holds} at the opposite limit $T\gg T^{\ast\ast}$. If the two CKCs are {\color{black} identical}, $T^{\ast\ast}=T^\ast$.

\subsubsection{$T\ll\left(\Gamma_{1},\:\Gamma_{2}\right)$,  Fermi-liquid regime:} \label{case4}
{\color{black} In the regime of vanishingly small temperatures $T\ll\left(\Gamma_{1},\:\Gamma_{2}\right)$}
the TP of the system behaves in accordance with the nonperturbative FL {\color{black} scenario}:  
\begin{eqnarray}
S & = & -\frac{{\color{black}12}\pi\gamma T}{7e}\left[\frac{|r_{1}|^{2}}{\Gamma_{1}}\ln\left(\frac{E_{C,1}}{\Gamma_{1}}\right)\sin\left(2\pi N_{1}\right)\right.\nonumber \\
 &  & \left.+\frac{|r_{2}|^{2}}{\Gamma_{2}}\ln\left(\frac{E_{C,2}}{\Gamma_{2}}\right)\sin\left(2\pi N_{2}\right)\right].
\end{eqnarray}
{\color{black} On the one hand,} the TP is a linear function of the temperature {\color{black} which is the hallmark for the FL regime. On the other hand,} the pre-factors are giant when both QDs are {\color{black} fine tuned by the gate voltages to the regime of} the vicinities of the Coulomb peaks. The system {\color{black} therefore is characterized by FL properties strongly renormalized by the scattering and interactions}. 
\begin{figure}
\begin{tabular}{c}
\vspace{-0.6cm}
\includegraphics[width=1\columnwidth]{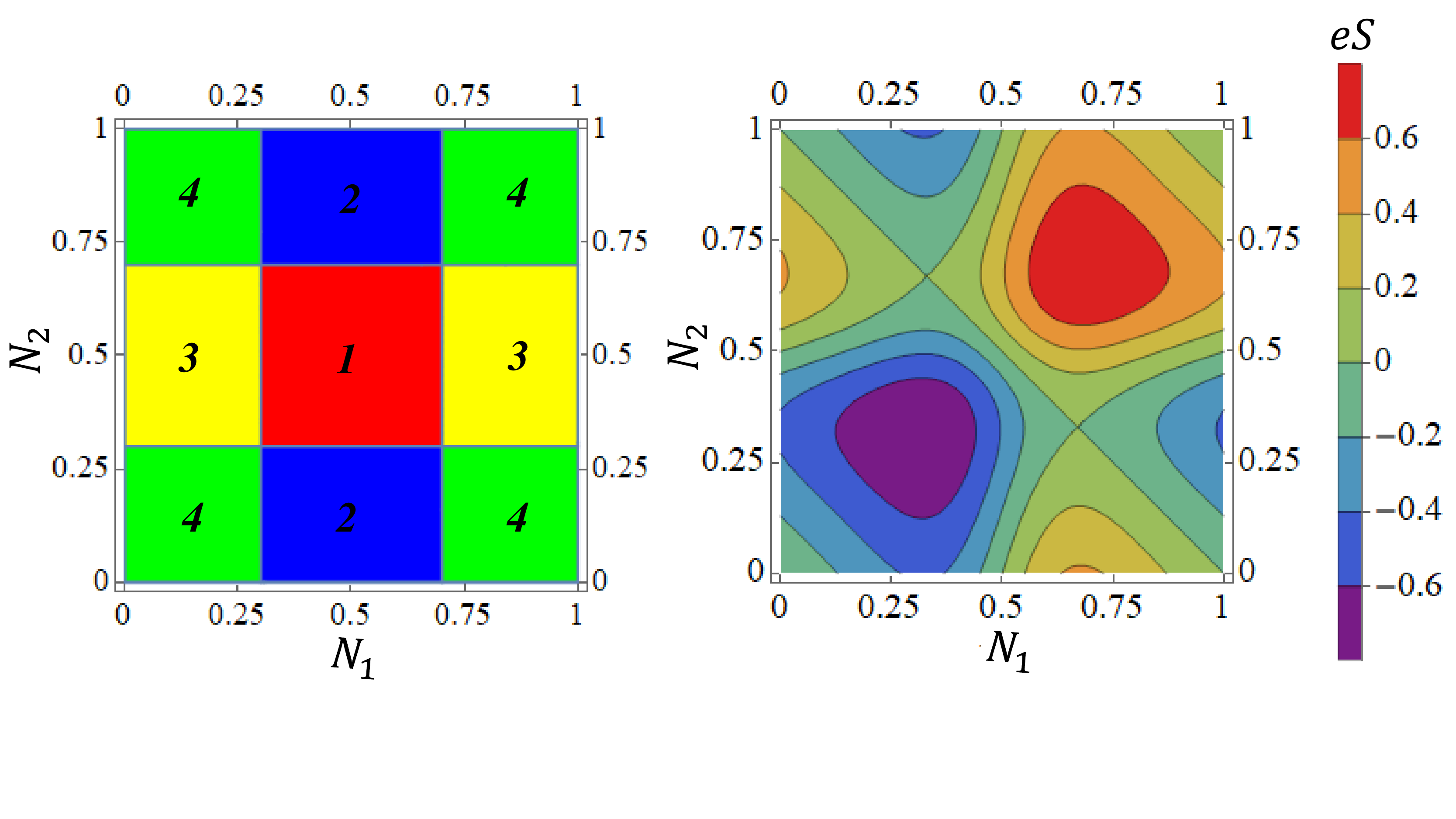}\tabularnewline
\end{tabular}
\vspace{-0.5cm}
\caption{Left panel: Region plot of 4 segments (the number marked on each one corresponds to the regime in the main text) as functions of the dimensionless gate voltages $N_1$ and $N_2$. 1-red region: $\Gamma_1,\:\Gamma_2\le T$, 2-blue region: $\Gamma_1\le T<\Gamma_2$, 3-yellow region: $\Gamma_2\le T<\Gamma_1$, 4-green region: $T<\Gamma_1,\:\Gamma_2$. Right panel: Contour plot of thermopower as function of  the dimensionless gate voltages $N_1$ and $N_2$ computed from general formula. Other parameters: $|r_1|^2=0.1$, $|r_2|^2=0.1$, $T/E_C=0.05$ ($E_{C,1}=E_{C,2}=E_C$).}
\label{fig_result1}
\end{figure} 
\begin{figure}
\begin{tabular}{c}
\includegraphics[width=1.5\columnwidth, angle=-90]{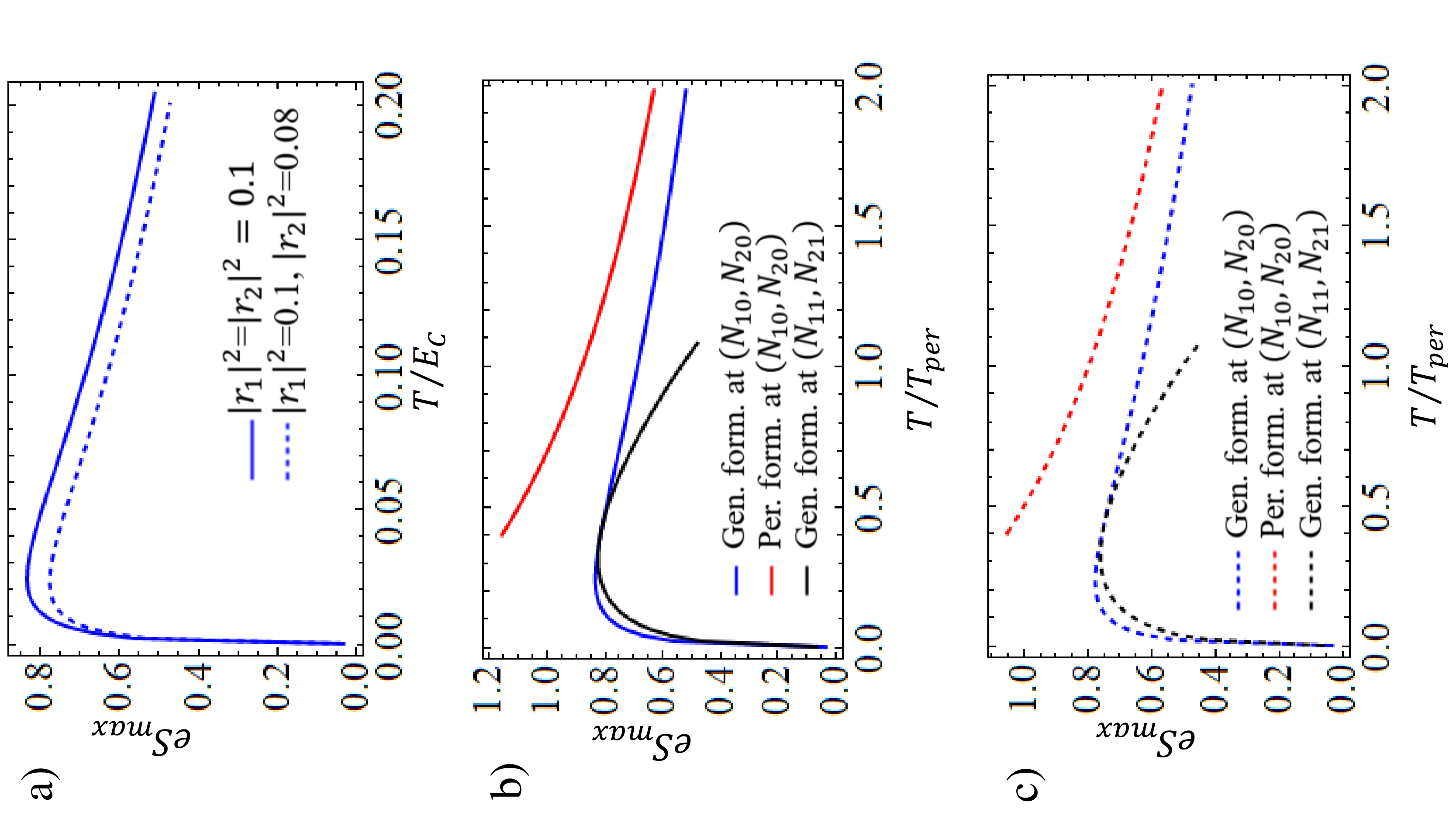}\tabularnewline
\end{tabular}
\caption{{\color{black} Maximum of thermopower as a function of temperature. The value $eS_{max}$ is achieved at $(N_{10},\,N_{20})$. a) the curves for plots are computed using general formulas [see Eqs. (35) and (38)] for different sets of reflection amplitudes: $|r_1|^2=|r_2|^2=0.1$ (continuous line) and $|r_1|^2=0.1,\,|r_2|^2=0.08$ (dashed line). b) and c) $eS_{max}$ as a function of $T/T_{per}$ ($T_{per}=\text{max}[|r_j|^2]E_C=0.1$ in both panels) are plot for these two parameter sets: blue lines are maximum of thermopower $eS_{max}$ taken from general formulas, red lines describe the thermopower computed from Eq. (\ref{eq:S_2NFL}) [perturbative solution] at $(N_{10},\,N_{20})$, while black lines correspond to the thermopower computed at the gate voltages $(N_{11},\,N_{21})$ at which $\Gamma_1(N_{11})=T$ and $\Gamma_2(N_{21})=T$   ($E_{C,1}=E_{C,2}=E_C$). The perturbative calculation's  validity  is defined by the condition $T/T_{\rm per}=1$ fulfilled on the right side of the plots.}}
\label{fig_result2}
\end{figure}
\begin{figure}
\begin{tabular}{c}
\includegraphics[width=1.5\columnwidth, angle=-90]{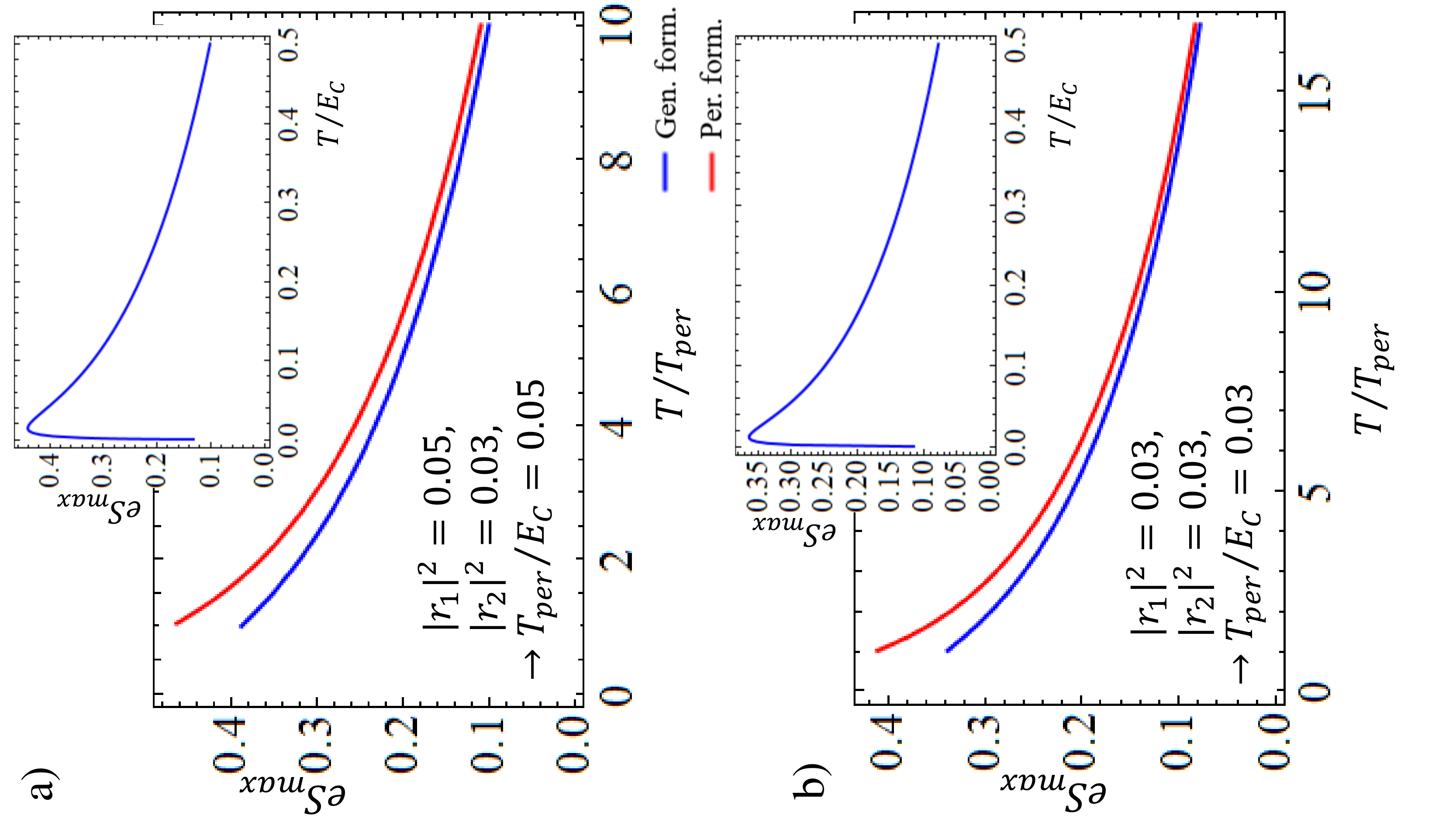}\tabularnewline
\end{tabular}
\vspace{-0.5cm}
\caption{{\color{black} Maximum of thermopower $eS_{\rm max}$ as a function of $T/T_{\rm per}$. Blue and red lines are computed from the general formulas and perturbation approach, respectively. a) $|r_1|^2=0.05,\,|r_2|^2=0.03$, so $T_{\rm per}/E_C=0.05$ and b) $|r_1|^2=|r_2|^2=0.03$, so $T_{\rm per}/E_C=0.03$. The inserts: Maximum of thermopower as a function of temperature. The perturbative result approaches the general one when $T/T_{\rm per}\gg 1$ ($E_{C,1}=E_{C,2}=E_C$).}}
\label{fig_result3}
\end{figure}

{\color{black}
The advantage of the nonperturbative solution in comparison to the perturbative one is that the latter is applicable only to the temperature regime $|r_j|^2E_{C,j}\ll T \ll E_{C,j}$ while the former can be used at any small temperature. With the appearance of the two Kondo resonance width energy scales $\Gamma_1, \: \Gamma_2$ in the nonperturbative treatment, we can {\color{black} identify} four different limit regimes as represented above. They {\color{black} correspond} to four regimes shown in Fig.\ref{fig_result1}-left panel as: the first - red, the second - blue, the third - yellow, and the fourth - green. The right panel of Fig.\ref{fig_result1} represents the contour plot of thermopower as a function of $(N_{1},\,N_{2})$. The perturbative result belongs to the first regime $\left(\Gamma_{1},\:\Gamma_{2}\right)\ll T$ in which the temperature scaling of TP behaves in accordance with the NFL scenario.  Thanks to the NFL enhancement of the heat proliferation the TP reaches the biggest value when $\left(\Gamma_{1},\:\Gamma_{2}\right)\rightarrow T$ at low temperature regime [see also in Fig. \ref{fig_result2}]. 

Depending on a domain of external parameters the TP of the 2SCKC behaves in accordance with either NFL or FL scenario. TP is a subject to one of the four regimes shown in Fig.\ref{fig_result1}-left panel when the temperature is varied.  Behavior of the TP maximum computed from the general formulas [see Eqs. (\ref{eq:genG}) and (\ref{eq:genGT})] as a function of temperature for the different sets of reflection amplitudes: $|r_1|^2=|r_2|^2=0.1$ (continuous line) and $|r_1|^2=0.1,\,|r_2|^2=0.08$ (dashed line) is shown on in Fig.\ref{fig_result2} a). In Fig.\ref{fig_result2} b) and c) we investigate these cases in details. Namely, the maximum of TP taken from general formulas and Eq. (\ref{eq:S_2NFL}) [perturbative solution] are plotted by the blue and red lines at the same dimensionless gate voltages $(N_{10},\,N_{20})$, respectively. Besides, in panels b) and c), the maximums of TP $eS_{\rm max}$ are plotted as functions of the ratio $T/T_{\rm per}$ where $T_{\rm per}$ is defined as $T_{\rm per}=\text{max}[|r_j|^2]E_C$. Thus, $T_{\rm per}=0.1$ in both panels. Clearly, the perturbative solution of the first regime ($\Gamma_1,\,\Gamma_2 \ll T$) is {\color{black} achieved} on the right side of the point at $T/T_{\rm per}=O(1)$. Fig.\ref{fig_result3} provides more details on the validity regime of the perturbative solution. Namely, the condition $T_{\rm per}\ll T\ll E_C$ can be adapted only when $|r_j|$ are small enough \cite{Thanh_VN_2}. 

On Fig.\ref{fig_result2} b) and c) we show by black lines the temperature behavior of the TP computed at specific values of dimensionless gate voltages $N_{11}$ and $N_{21}$ {\color{black} satisfying conditions} $\Gamma_1(N_{11})=T$ and $\Gamma_2(N_{21})=T$. Remarkably, the regimes where TP gets maximum value, are at $(\Gamma_1,\Gamma_2)\sim T$. On the one hand, 
$(\Gamma_1,\Gamma_2)>T$ at very low temperatures (the left side of the crossing point between the blue and black lines). On the other hand, $(\Gamma_1,\Gamma_2) < T$ at higher temperatures (the right side of the crossing point). We conclude that to observe the fingerprints of the NFL behavior in the thermoelectric transport of the 2SCKCs it is sufficient to perform several sets of experiments at different temperatures from the highest one $T\gg |r_j|^2E_{C,j}$ to the lowest possible
(notice that the temperature is anyway bounded from below by the mean level spacing of the QD being the smallest energy parameter of the model). Besides,
the condition of the vicinity to the Coulomb peaks provides an access to crossing three distinct regimes: either ($1\rightarrow 2 \rightarrow 4$) or ($1\rightarrow 3 \rightarrow 4$) where the crossover between NFL and FL behavior is predicted to be pronounced (here digits $n=1\to 4$ refer to four distinct temperature regimes discussed in Section IV and analysed in gross details in subsections B.n. Necessary details of calculations are presented in the Appendix.)
}

{\color{black}\section{Discussion and conclusion}}

\begin{figure}
\begin{tabular}{c}
\includegraphics[width=1\columnwidth]{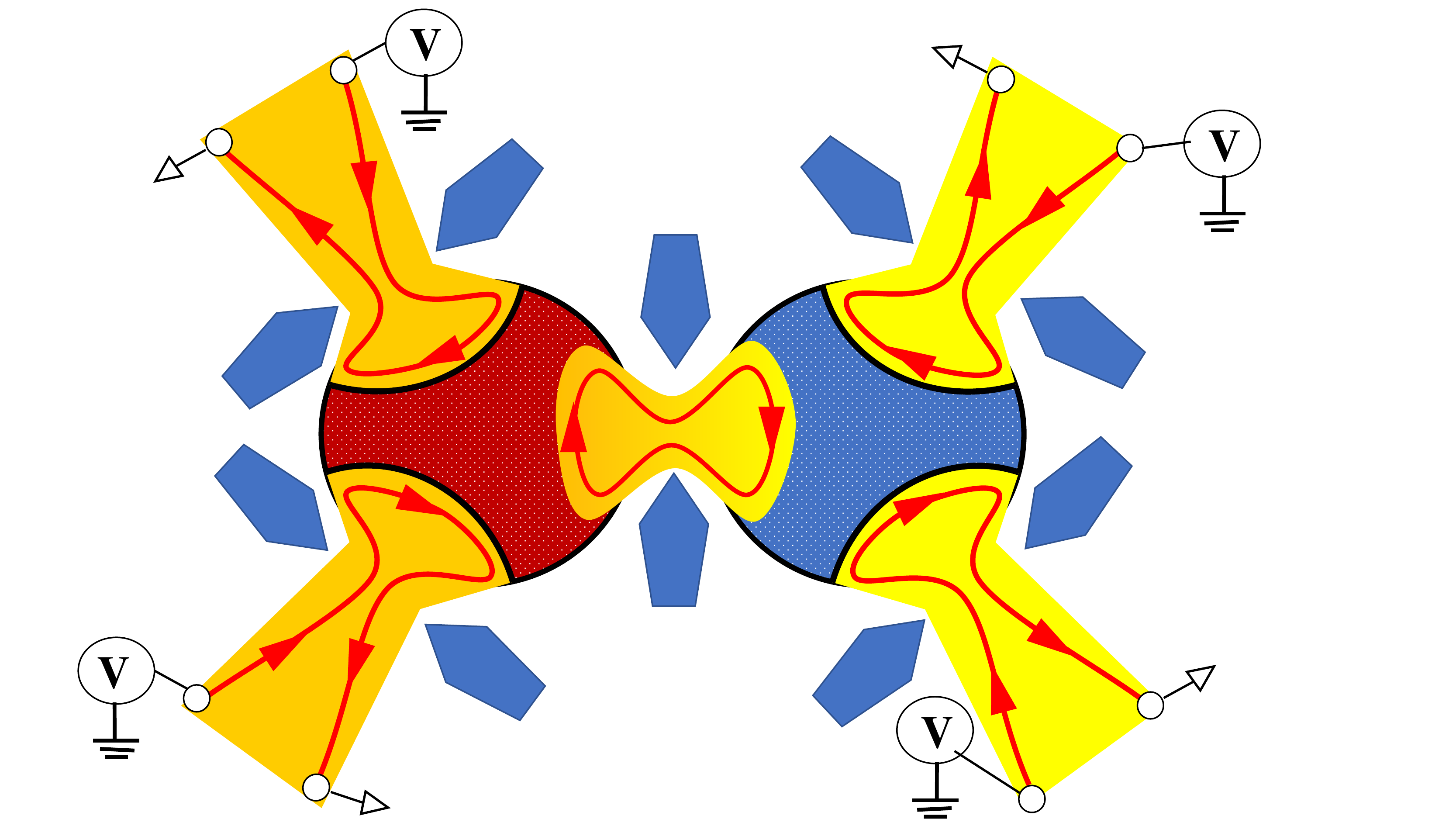}\tabularnewline
\end{tabular}\caption{Schematic of a strong coupling between two charge Kondo circuits. Notations are the same as on Figure \ref{Fig1} with two distinct differences:
i) different color scheme used for the left (orange, hot) and right (yellow, cold) parts of the circuit; ii) the central part of the system is modified. The tunnel contact in the center of Fig \ref{Fig1} is replaced by almost transparent QPC. The chiral edge mode strongly couples QD1 and QD2. The temperature gradient (shown by continuous change of colour from orange to yellow) is applied across the central QPC.}
\label{Fig2}
\end{figure}
{\color{black}
The investigation of TP for the weak coupling between two CKCs in both cases: 1CK - 2CK and 2CK - 2CK shows the competition between the FL and NFL picture. However, the windows of parameters to observe the FL property are much broader than the windows to access the NFL one. The reason is that the NFL intermediate coupling fixed points of MCK are hyperbolic and therefore unstable.
The results of this work not only cover the perturbative accessible regimes, which have been represented in Ref. \cite{thanh2018}, but also show a rich property of the TP in different domains of  parameters.

{\color{black} \textit{The figure of merit $ZT$:} 
In order to estimate the figure of merit $ZT$ in the 2SCKCs we re-write the Eq (\ref{ZTeq}) in the following form:
\begin{equation}
ZT=\frac{S^2}{L_0\cdot R(T)},\;\;\;\; \frac{\cal K}{GT}=L_0\cdot R(T)
=L_0\cdot R_L -S^2
\end{equation}
On the one hand, the $ZT$ is proportional to the second power of the TP. On the other hand, it is inversely proportional to the Lorenz Ratio $R_L$  \cite{Kis2023}. The conventional WF law predicts $R_L$$=$$1$ at $T$$=$$0$ with vanishingly small corrections at low temperatures. The generalized WF law discussed in  \cite{Kis2023} says that the Lorenz Ratio in the 2SCKCs is in general greater than unity and in particular, bounded from above by the value 
$R^{\rm max}_L$$=$$27$$/$$7$. Therefore, two competing effects take place: the NFL scenario significantly enhances the thermoelectric power at low temperatures compare to the FL value while Anderson orthogonality catastrophe results in increase of the Lorenz Ratio  \cite{Kis2023}. Yet another ``bad news" for the figure of merit are associated with vanishing of TP at the low temperature regime and vanishing of TP for the almost transparent QPC setups (small reflection amplitude is  a PH breaking parameter which has to be controllable small for the validity of the whole theory).  As one can see it from Fig.4 and Fig. 5, the maximum of TP is reached at the crossover temperature regime 
$\left(\Gamma_{1},\:\Gamma_{2}\right)\approx T$ at the very vicinity of the Coulomb peaks $N_j\approx 0.5\pm\delta N_j$. Following Matveev-Andreev \cite{andreevmatveev}  we estimate $\delta N_j\sim 1/|r_j| \sqrt{T/E_{C,j}}$.  It finally results in the ``best" $ZT$ estimate
\begin{equation}
[ZT]_{\rm max} \to  \frac{a}{R_L} |r|^2\frac{\Gamma}{E_C}\ln^2\left(\frac{E_C}{\Gamma}\right)
\end{equation}
with $a \lesssim 1$.  For this estimation we choose $|r_1|=|r_2|=|r|$, $\Gamma_1=\Gamma_2=\Gamma$ and $E_{C,1}=E_{C,2}=E_C$.  For realistic 2SCKCs 
$[ZT]_{\rm max}\sim |r|^4\ln^2(|r|^2)\ll 1$.
}    

Extending the proposal of the weak coupling between two CKCs \cite{thanh2018} to the regime of almost transparent QPC in the central area of the {\color{black}2SCKC}, the very recent experiment \cite{Gordon2023} and theory \cite{Karki2022,Z3_DCK} have investigated the strong coupling limit. Let us comment on the connection between the weak and strong coupling regimes of the {\color{black}2SCKCs}. In Ref. \cite{thanh2018} we have considered the {\color{black}2SCKC} weakly connecting 1CK-1CK or 1CK-2CK or 2CK-2CK. The
same realization for the strong coupling of two Kondo
simulators has also been theoretically suggested in \cite{thanh2018}  and  experimentally realized recently in \cite{Gordon2023} for 1CK-1CK coupling \cite{com2}. One of the most exciting theoretical predictions of the  two-impurity single channel Kondo  effect \cite{Jones_Varma,Gan_95,Logan2012} is a possibility to map the model under certain assumptions onto the 2CK Hamiltonian.
Interestingly, the Refs. \cite{Gordon2023,Karki2022,Z3_DCK} showed that at the triple degeneracy point of the {\color{black}2SCKC} $Z_3$ symmetry and corresponding local parafermion emerge. It is straightforward to extend the idea \cite{thanh2018} to MCK-NCK strong coupling (see Fig \ref{Fig2}). Suppose that there are $M>1$ identical QPCs in the left hand side of the {\color{black}2SCKC} and $N>1$ identical QPCs in the right side of it. The total degeneracy is $M+1+N $ and corresponding emergent local symmetry is $Z_{M+N+1}$. There are three important $(M,N)$ realizations accessible through existing experimental setups: i) $(2,1)$ or $(1,2)$ connecting 1CK and 2CK with emergent symmetry $Z_4$; ii) $(2,2)$ and iii) $(3,1)$ or $(1,3)$ with emergent symmetry $Z_5$.
Corresponding weak link setups are characterized by the symmetries: i) $U(1)\times Z_2$;
ii) $Z_2\times Z_2$ and iii) $U(1)\times Z_3$.  As the weak coupling regimes of ii) and iii) are clearly distinct, being characterized by both different symmetries and different  Lorenz ratios (see Ref. \cite{Kis2023} for more details), it is interesting to examine regimes ii) and iii) in the strong coupling limit. In particular it is important to understand the symmetry of {\it local} parafermion emerging in the strong link setup. In addition, switching between different intermediate coupling fixed points results in crossovers between various fractionalized modes manifesting itself in distinctly different regimes of the charge and heat transport. 

The weak link regime discussed in this manuscript was analysed using a standard approach based on the transport integrals \cite{thanh2018}.
The validity of this approach is justified by an assumption that both temperature and voltage drops occur exactly {\color{black} at} the central tunnel barrier. As a result, both the left and the right parts of the {\color{black}2SCKC} are considered at thermal and mechanical equilibrium being characterized by certain temperature $T$ and chemical potential $\mu$. This approach is clearly invalid for the strong link between two sides of the Kondo simulator where both the temperature and the voltage changes continuously across the central QPC. The full-fledged linear response theory of the charge and heat transport  across the strong link of the two-site Kondo simulators can be constructed by using {\it Luttinger's  pseudo-gravitational approach}  \cite{lut1,lut2} or {\it thermo-mechanical} potential \cite{lut3,lut4} method in combination with Kubo equations. The theory beyond linear response requires also using Keldysh formalism \cite{lut3,lut4} and represents an interesting and important direction for the future investigation.

In {\color{black}summary}, we revisited the thermoelectric transport at the weak link of the {\color{black}2SCKC} model proposed in the Ref. \cite{thanh2018}. The Abelian bosonization approach is used for both 1CK and 2CK setup while the refermionization technique is applied in order to solve the 2CK model nonperturbatively. We show the different windows of the parameter set where the {temperature dependence of} TP behaves either the full FL or NFL characteristics or the competition between these properties. The nonperturbative results not only cover the perturbative results but also be applicable in the lower temperature regime $T<|r_j|^2E_{C,j}$. We predict that the TP is enhanced in the {\color{black}2SCKC} in comparison with the single CKC setup. 
{\color{black} Indeed, complex charge Kondo circuits showing the diversity of the competition between the FL and NFL properties represent new classes of  the quantum simulators characterized by a significant improvement of the thermoelectric properties. The models  describing the charge and heat transport through these devices deserve extensive theoretical and experimental studies in the future. The realistic figures of merit of the {\color{black}2SCKCs} are, however, still far from approaching the numbers demanded by the quantum computer applications.  Detailed investigation of the multi-site quantum simulators in the NFL operational regimes provides a promising avenue for the new quantum technologies. Finally}, we propose to use the experimental implementation in Ref. \cite{Gordon2023} for investigating the different parafermion contributions to the quantum thermoelectricity when the coupling between QDs is switched from weak to strong.
\bigskip\\
\section*{Acknowledgement}
This research in Hanoi is funded by Vietnam Academy of Science and
Technology (program for Physics development) under grant number KHCBVL.06/23-24. The work of M.N.K is conducted within the framework of the Trieste Institute for Theoretical Quantum Technologies (TQT). M.N.K also acknowledges the support from the Alexander von Humboldt Foundation for the research visit to IFW Dresden. {\color{black}The research of M.N.K. was supported in part by the National Science Foundation under Grants No. NSF PHY-1748958 and PHY-2309135.  M.N.K.  acknowledges support of the Institute Henri Poincare (UAR 839 CNRS-Sorbonne Universite) and LabEx CARMIN (ANR-10LABX-59-01).}

\section*{Appendix}

In this Appendix we represent the details of the different approaches at different limits in order to obtain the results shown in the Subsection \ref{2CK_2CK}. 

1, If $p_{1}\rightarrow0,p_{2}\rightarrow0$, we have: 
\begin{align}
\lim_{p_{1}\rightarrow0}\frac{p_{1}}{\left(z+\frac{u}{2}\right)^{2}+p_{1}^{2}} & =\pi\delta\left(z+\frac{u}{2}\right),\\
\lim_{p_{2}\rightarrow0}\frac{p_{2}}{\left(z-\frac{u}{2}\right)^{2}+p_{2}^{2}} & =\pi\delta\left(z-\frac{u}{2}\right).
\end{align}
As a result
\begin{eqnarray}
 &  & \lim_{p_{2}\rightarrow0}\frac{F_{T,m}\left(p_{1}\rightarrow0,p_{2}\right)}{p_{2}}\nonumber \\
 &  & =\lim_{p_{2}\rightarrow0}\int_{-\infty}^{\infty}du\frac{\pi u^{3}\left[u^{2}+4\pi^{2}\right]}{{\color{black}2}\sinh\left[u\right]\left[u^{2}+p_{2}^{2}\right]}\nonumber \\
 &  & =\int_{-\infty}^{\infty}du\frac{\pi u\left[u^{2}+4\pi^{2}\right]}{{\color{black}2}\sinh\left[u\right]}=\frac{{\color{black}9}\pi^{5}}{{\color{black}8}},
\end{eqnarray}
and, finally
\begin{equation}
\lim_{p_{1}\rightarrow0}\frac{F_{T,s}\left(p_{1},p_{2}\rightarrow0\right)}{p_{1}}=\frac{{\color{black}9}\pi^{5}}{{\color{black}8}}~.
\end{equation}
We obtain the electric conductance as 
\begin{align}
G & =\frac{{\color{black}\pi^4}G_{C}T^2}{6\gamma^2 E_{C,1}E_{C,2}},
\end{align}
}
and the thermoelectric coefficient as 
\begin{eqnarray}
 &  & \!\!\!\!G_{T}=-\frac{{\color{black}3}\pi^{4}G_{C}T^{2}}{{\color{black}16}e\gamma E_{C,1}E_{C,2}}\left[|r_{1}|^{2}\ln\left(\frac{E_{C,1}}{T}\right)\sin\left(2\pi N_{1}\right)\right.\nonumber \\
 &  & \left.+|r_{2}|^{2}\ln\left(\frac{E_{C,2}}{T}\right)\sin\left(2\pi N_{2}\right)\right].
\end{eqnarray}

2, If $p_{1}\rightarrow0,p_{2}\gg1$ we have: 
\begin{eqnarray}
 &  & F_{C}\left(p_{1}\rightarrow0,p_{2}\right)=\int_{-\infty}^{\infty}du\frac{\pi p_{2}u\left[u^{2}+4\pi^{2}\right]}{\sinh\left[u\right]\left[u^{2}+p_{2}^{2}\right]},\nonumber \\
 &  & \!\!\!\!\!\!\!\!F_{C}\left(p_{1}\rightarrow0,p_{2}\gg1\right)\!=\!\frac{\pi}{p_{2}}\!\!\int_{-\infty}^{\infty}\!\!\!\!\!\!du\frac{u\left[u^{2}+4\pi^{2}\right]}{\sinh\left[u\right]}\!=\!\frac{9\pi^{5}}{4p_2},
\end{eqnarray}
and
\begin{align}
\!\!\!\!\!\!\!\!F_{T,m}\left(p_{1}\rightarrow0,p_{2}\gg1\right)\!=\!\frac{\pi}{{\color{black}2}p_2}\!\!\int_{-\infty}^{\infty}\!\!\!\!\!\!\!du\frac{u^{3}\left[u^{2}+4\pi^{2}\right]}{\sinh\left[u\right]}\!=\!\frac{3\pi^{7}}{{\color{black}4}p_2}.
\end{align}
The calculation of the $F_{T,s}$ is a bit complicated, which concerns the principal value (PV) as follows.
\begin{align}
 & \!\!\!\!\frac{F_{T,s}\left(p_{1}\rightarrow0,p_{2}\gg1\right)}{p_{1}}=\frac{1}{p_{2}}\text{PV}\int_{-\infty}^{\infty}\!\!\!\!dz\int_{-\infty}^{\infty}\!\!\!\!du\nonumber \\
 & \!\!\!\!\times\frac{uz\left[u^{2}+4\pi^{2}\right]}{\left(z+\frac{u}{2}\right)\sinh\left(\frac{u}{2}\right)\left[\cosh\left(z\right)+\cosh\left(\frac{u}{2}\right)\right]}\nonumber \\
 & \!\!\!\!\!\!=\frac{8}{p_{2}}\!\!\int_{-\infty}^{\infty}\!\!\!\!dp\!\int_{-\infty}^{\infty}\!\!\!\!dq\left\{\frac{q\left[q^{2}+{\color{black}\pi^2}\right]}{\sinh\left(q\right)\cosh\left(p/2\right)\cosh\left(p/2-q\right)}\right.\nonumber \\
 & \!\!\!\!\left.-\frac{\tanh\left(p/2\right)q^{2}\left[q^{2}+{\color{black}\pi^2}\right]}{p\cosh\left(p/2+q\right)\cosh\left(p/2-q\right)}\right\} \nonumber\\
 &=\frac{192\pi^4}{25p_2}.
\end{align}
The electric conductance $G$ and the thermoelectric coefficient $G_T$ is computed at the first non-zero term in the nonperturbative treatment are 
\begin{equation}
G=\frac{3G_{C}\pi^{5}T^{3}}{32\gamma^{2}\Gamma_{2}E_{C,1}E_{C,2}},
\end{equation}
\begin{eqnarray}
& & \!\!\!\!\!\!G_{T}=-\frac{{\color{black}32}G_{C}\pi^3T^3}{25e\gamma E_{C,1}E_{C,2}\Gamma_2}\left[|r_1|^2\ln\left(\frac{E_{C,1}}{T}\right)\sin\left(2\pi N_1\right)\right.\nonumber \\
& &\!\!\!\!\!\!\!\!\! \left.+\frac{25\pi^3}{{\color{black}256}}|r_2|^2\frac{T}{\Gamma_2}\ln\left(\frac{E_{C,2}}{\Gamma_2}\right)\sin\left(2\pi N_2\right)\right].
\end{eqnarray}

3, $p_{1}\gg1,p_{2}\rightarrow0$: This limit is opposite to the second limit. The calculation process is the same as the above one.
\begin{align}
 & F_{C}\left(p_{1}\gg1,p_{2}\rightarrow0\right)\nonumber \\
 & =\frac{\pi}{p_{1}}\int_{-\infty}^{\infty}du\frac{u\left[u^{2}+4\pi^{2}\right]}{\sinh\left[u\right]}=\frac{9\pi^{5}}{4p_{1}},
\end{align}
\begin{eqnarray}
&& \frac{F_{T,m}\left(p_{1}\gg1,p_{2}\rightarrow0\right)}{p_{2}}=\frac{1}{p_{1}}\text{PV}\int_{-\infty}^{\infty}dz\int_{-\infty}^{\infty}du\nonumber \\
&& \times\frac{uz\left[u^{2}+4\pi^{2}\right]}{\left(z-\frac{u}{2}\right)\sinh\left(\frac{u}{2}\right)\left[\cosh\left(z\right)+\cosh\left(\frac{u}{2}\right)\right]}\nonumber\\
&&=\frac{192\pi^4}{25p_1},
\end{eqnarray}
\begin{equation}
F_{T,s}\left(p_{1}\gg1,p_{2}\rightarrow0\right)=\frac{3\pi^{7}}{{\color{black}4}p_1}.    
\end{equation}
The electric conductance is 
\begin{equation}
G=\frac{3G_{C}\pi^{5}T^{3}}{32\gamma^{2}E_{C,1}E_{C,2}\Gamma_{1}},
\end{equation}
and the thermoelectric coefficient is 
\begin{eqnarray}
& & \!\!\!\!\!\!G_{T}=-\frac{{\color{black}32}G_{C}\pi^3T^3}{25e\gamma E_{C,1}E_{C,2}\Gamma_1}\left[|r_2|^2\ln\left(\frac{E_{C,2}}{T}\right)\sin\left(2\pi N_2\right)\right.\nonumber \\
& &\!\!\!\!\!\!\!\!\! \left.+\frac{25\pi^3}{{\color{black}256}}|r_1|^2\frac{T}{\Gamma_1}\ln\left(\frac{E_{C,1}}{\Gamma_1}\right)\sin\left(2\pi N_1\right)\right].
\end{eqnarray}

4, If $p_{1}\gg1,p_{2}\gg1$, we simply remove the terms which are summed with $p_1^2$ and $p_2^2$ in the denominator of the formula (\ref{FFF}). We then obtain: 
\begin{align}
 & F_{C}\left(p_{1}\gg1,p_{2}\gg1\right)=\frac{1}{p_{1}p_{2}}\int_{-\infty}^{\infty}dz\int_{-\infty}^{\infty}du\nonumber \\
 & \times\frac{u\left[u^{2}+4\pi^{2}\right]}{\sinh\left[\frac{u}{2}\right]\left[\cosh\left(z\right)+\cosh\left(\frac{u}{2}\right)\right]}=\frac{64\pi^{4}}{5p_{1}p_{2}}.
\end{align}
\begin{align}
 & F_{T,m}\left(p_{1}\gg1,p_{2}\gg1\right)=\frac{1}{p_{1}p_{2}}\int_{-\infty}^{\infty}dz\int_{-\infty}^{\infty}du\nonumber \\
 & \times\frac{\left(z-\frac{u}{2}\right)uz\left[u^{2}+4\pi^{2}\right]}{\sinh\left[\frac{u}{2}\right]\left[\cosh\left(z\right)+\cosh\left(\frac{u}{2}\right)\right]}=\frac{192\pi^{6}}{35p_{1}p_{2}}.
\end{align}
\begin{align}
 & F_{T,s}\left(p_{1}\gg1,p_{2}\gg1\right)=\frac{1}{p_{1}p_{2}}\int_{-\infty}^{\infty}dz\int_{-\infty}^{\infty}du\nonumber \\
 & \times\frac{\left(z+\frac{u}{2}\right)uz\left[u^{2}+4\pi^{2}\right]}{\sinh\left[\frac{u}{2}\right]\left[\cosh\left(z\right)+\cosh\left(\frac{u}{2}\right)\right]}=\frac{192\pi^{6}}{35p_{1}p_{2}}.
\end{align}
The electric conductance and the thermoelectric coefficient in this
limit are 
\begin{equation}
G=\frac{8G_{C}\pi^{4}T^{4}}{15\gamma^{2}E_{C,1}E_{C,2}\Gamma_{1}\Gamma_{2}},
\end{equation}
\begin{eqnarray}
 &  & G_{T}=-\frac{{\color{black}32}G_{C}\pi^5T^5}{{\color{black}35}e\gamma E_{C,1}E_{C,2}\Gamma_{1}\Gamma_{2}}\nonumber \\
 &  & \times\left\{ \frac{|r_{1}|^{2}}{\Gamma_{1}}\ln\left(\frac{E_{C,1}}{T+\Gamma_{1}}\right)\sin\left(2\pi N_{1}\right)\right.\nonumber \\
 &  & +\left.\frac{|r_{2}|^{2}}{\Gamma_{2}}\ln\left(\frac{E_{C,2}}{T+\Gamma_{2}}\right)\sin\left(2\pi N_{2}\right)\right\}. 
\end{eqnarray}

\end{document}